\documentclass[aps,pra,amsmath,amssymb,tightenlines,epsfig,floatfix,twocolumn,superscriptaddress, longbibliography]{revtex4-1}
\usepackage{graphicx}
\usepackage{dcolumn}	
\usepackage{bm}			
\usepackage{amsfonts}
\usepackage{xspace}
\usepackage{color}
\usepackage{epstopdf}
\usepackage{multirow}

\usepackage{mathtools}

\DeclarePairedDelimiter\ket{\lvert}{\rangle}

\begin{document}

\newcommand{\psihat}{\ensuremath{\hat{\psi}}\xspace}
\newcommand{\psihatd}{\ensuremath{\hat{\psi}^{\dagger}}\xspace}
\newcommand{\ahat}{\ensuremath{\hat{a}}\xspace}
\newcommand{\Ham}{\ensuremath{\mathcal{H}}\xspace}
\newcommand{\ahatd}{\ensuremath{\hat{a}^{\dagger}}\xspace}
\newcommand{\bhat}{\ensuremath{\hat{b}}\xspace}
\newcommand{\bhatd}{\ensuremath{\hat{b}^{\dagger}}\xspace}
\newcommand{\boldr}{\ensuremath{\mathbf{r}}\xspace}
\newcommand{\dr}{\ensuremath{\,d^3\mathbf{r}}\xspace}
\newcommand{\tr}{\ensuremath{\,\mathrm{Tr}}\xspace}
\newcommand{\dk}{\ensuremath{\,d^3\mathbf{k}}\xspace}
\newcommand{\etal}{\emph{et al.\/}\xspace}
\newcommand{\ie}{i.e.\ }
\newcommand{\eq}[1]{Eq.\ (\ref{#1})\xspace}
\newcommand{\fig}[1]{Fig.\ \ref{#1}\xspace}
\newcommand{\abs}[1]{\left| #1 \right|}
\newcommand{\proj}[2]{\left| #1 \rangle\langle #2\right| \xspace}
\newcommand{\Qhat}{\ensuremath{\hat{Q}}\xspace}
\newcommand{\Qhatd}{\ensuremath{\hat{Q}^\dag}\xspace}
\newcommand{\phihatd}{\ensuremath{\hat{\phi}^{\dagger}}\xspace}
\newcommand{\phihat}{\ensuremath{\hat{\phi}}\xspace}
\newcommand{\boldk}{\ensuremath{\mathbf{k}}\xspace}
\newcommand{\boldp}{\ensuremath{\mathbf{p}}\xspace}
\newcommand{\boldsigma}{\ensuremath{\boldsymbol\sigma}\xspace}
\newcommand{\boldalpha}{\ensuremath{\boldsymbol\alpha}\xspace}
\newcommand{\grad}{\ensuremath{\boldsymbol\nabla}\xspace}
\newcommand{\parti}[2]{\frac{ \partial #1}{\partial #2} \xspace}
 \newcommand{\vs}[1]{\ensuremath{\boldsymbol{#1}}\xspace}
\renewcommand{\v}[1]{\ensuremath{\mathbf{#1}}\xspace}
\newcommand{\Psihat}{\ensuremath{\hat{\Psi}}\xspace}
\newcommand{\Psihatd}{\ensuremath{\hat{\Psi}^{\dagger}}\xspace}
\newcommand{\Vhatd}{\ensuremath{\hat{V}^{\dagger}}\xspace}
\newcommand{\Xhat}{\ensuremath{\hat{X}}\xspace}
\newcommand{\Xhatd}{\ensuremath{\hat{X}^{\dag}}\xspace}
\newcommand{\Yhat}{\ensuremath{\hat{Y}}\xspace}
\newcommand{\Jhat}{\ensuremath{\hat{J}}\xspace}
\newcommand{\Yhatd}{\ensuremath{\hat{Y}^{\dag}}\xspace}
\newcommand{\Uhat}{\ensuremath{\hat{U}^{\dag}}\xspace}
\newcommand{\jhat}{\ensuremath{\hat{J}}\xspace}
\newcommand{\lhat}{\ensuremath{\hat{L}}\xspace}
\newcommand{\Nhat}{\ensuremath{\hat{N}}\xspace}
\newcommand{\rhohat}{\ensuremath{\hat{\rho}}\xspace}
\newcommand{\ddt}{\ensuremath{\frac{d}{dt}}\xspace}
\newcommand{\nset}{\ensuremath{n_1, n_2,\dots, n_k}\xspace}
\newcommand{\Var}{\ensuremath{\mathrm{Var}}\xspace}
\newcommand{\Erf}{\ensuremath{\mathrm{Erf}}\xspace}

\newcommand{\notes}[1]{{\color{purple}#1}}
\newcommand{\sah}[1]{{\color{blue}#1}}
\newcommand{\spn}[1]{{\color{magenta}#1}}



\title{Generating Macroscopic Superpositions with Interacting Bose-Einstein Condensates: Multi-Mode Speed-Ups and Speed Limits.}

\author{Samuel P. Nolan} 
\affiliation{School of Mathematics and Physics, The University of Queensland, Brisbane, Queensland, Australia}
\email{samuel.nolan@uqconnect.edu.au}
\author{Simon A. Haine}
\affiliation{Department of Quantum Science, Australian National University, Canberra, Australia}
\affiliation{Department of Physics and Astronomy, University of Sussex, Brighton, United Kingdom}

\date{\today}

\begin{abstract}

We theoretically investigate the effect of multi-mode dynamics on the creation of macroscopic superposition states (spin-cat states) in Bose-Einstein condensates via one-axis twisting. A two-component Bose-Einstein condensate naturally realises an effective one-axis twisting interaction, under which an initially separable state will evolve toward a spin-cat state. However, the large evolution times necessary to realise these states is beyond the scope of current experiments. This evolution time is proportional to the degree of asymmetry in the relative scattering lengths of the system, which results in the following trade-off; faster evolution times are associated with an increase in multi-mode dynamics, and we find that generally multi-mode dynamics reduce the degree of entanglement present in the final state. However, we find that highly entangled cat-like states are still possible in the presence of significant multi-mode dynamics, and that these dynamics impose a speed-limit on the evolution such states.
\end{abstract}

\maketitle

\section{Introduction} 
Atom interferometers are precision measurement devices with many applications in both fundamental science and industry \cite{Cronin2009}. Aside from a handful of proof-of-principle experiments, the phase sensitivity $\Delta \phi$ of most atom interferometers with $N$ atoms is shot-noise limited, $\Delta \phi \geq 1/\sqrt{N}$ \cite{Pezze:2016_review}. This precision limit may be surpassed by employing states that exhibit $N$-body entanglement \cite{Giovannetti2006, Pezze2009}, up to the ultimate Heisenberg limit $\Delta \phi \geq 1/N$. The states that achieve this maximum sensitivity are the `spin-cat states', which are coherent macroscopic superpositions of the maximum and minimum projections of the collective spin \cite{Bollinger1996}. As well as providing Heisenberg limited sensitivity \cite{Bollinger1996, Pezze:2007b, Haine2015a, Pezze:2016_review}, it has recently been shown that the ability to create these states can provide robustness against detection noise \cite{Nolan2017b, Fang2017, Huang:2018b, Haine:2018b}. These states can be generated from unentangled states via one-axis twisting (OAT) dynamics \cite{Agarwal1997, Molmer1999}. One-axis twisting (OAT) \cite{Kitagawa1993} is naturally realised due to atom-atom interactions in two-component Bose-Einstein condensates (BECs) \cite{Sorensen2002a}, and has emerged as an extremely successful method of generating entanglement in BECs \cite{Esteve2008, Riedel2010, Gross2010, Berrada2013, Ockeloen2013, Schmied2016}. OAT dynamics have also been demonstrated with cold atoms in a cavity-QED setting \cite{SchleierSmith2010b, Leroux2010, Leroux2010b, Leroux2012, Hosten2016b} and with trapped ions \cite{Meyer2001, Leibfried2004, Liebfried2005, Monz2011, Bohnet2016}. Current OAT experiments are performed with small, tightly confined condensates \cite{Pezze:2016_review}. In this regime spatial dynamics are unimportant and may be neglected, resulting in a single-mode analysis. Usually, the twisting rate is slow relative to timescales associated with sources of decoherence such as dephasing and particle losses, and thus making spin-cat states with OAT is outside the realm of current experiments \cite{Strobel2014, Pezze:2016_review}. This challenge is compounded by the notorious fragility of these states \cite{Aolita2008, DemkowiczDobrzanski2012, Huang2015a, Modi2016, Nolan2017, Lopez_Incera:2018, Frowis:2018_review}. Nevertheless, small spin-cat states have been created in other systems, such as superconducting flux qubits \cite{Friedman2000}, nuclear spins \cite{Jones2009}, angular momentum states of a single Rydberg atom \cite{Facon2016} and in trapped ions \cite{Leibfried2004, Liebfried2005, Monz2011}. Macroscopic superpositions of optical coherent states have also been realised \cite{Grangier2007}. 

More rapid twisting dynamics occur in systems with highly asymmetric scattering lengths \cite{Li2009, Gross2010, Riedel2010}, which usually results in significant multimode dynamics, especially when combined with large particle number \cite{Li2009, Haine2009}. 
In Ref. \cite{Pawlowski2017} the authors perform a multi-mode analysis of spin-cat states, with a focus on studying losses and finite temperature effects as sources of decoherence. In this paper we take a slightly different approach, and study the effect of multi-mode dynamics on OAT with the goal of producing spin-cat states (perhaps approximately) more rapidly than in a single-mode regime. This approach has already been suggested as a possibility for enhancing spin-squeezing under OAT \cite{Haine2014, Laudat:2018}. We do not study decoherence \textit{per se} as the state remains pure, however multi-mode dynamics can take the system away from ideal OAT behaviour and thus, compared to single-mode dynamics, may reduce the entanglement of the final state. 

The structure of this paper is as follows: In Section \ref{sec:SM} we revise ideal single-mode OAT and spin-cat states. Starting from the general multi-mode Hamiltonian for a two-component BEC, we show that under a single-mode approximation the dynamics reduce to an effective OAT interaction which can be used to generate nonclassical states. We define spin-cat states, as well as the quantum Fisher information (QFI), which we use throughout this paper to quantify the metrological usefulness of states produced under multi-mode OAT. 
In Section \ref{sec:castinsinatra} we argue that working in a more multi-mode regime should give rise to faster twisting dynamics, and introduce a numerical formalism that we use throughout the remainder of the paper. In Section \ref{sec:MMQFI}, we investigate the effect of multi-mode dynamics on the QFI. As these states are no longer Heisenberg limited, strictly speaking they are not spin-cat states. Nevertheless, we wish to investigate conditions under which large QFI states may be created which would still be extremely valuable resources for quantum-enhanced metrology. Thus, in Section \ref{sec:MMcats} we explore a range of parameters and find that so long as the chemical potential is carefully chosen, states with large QFI are still achievable in a highly multi-mode regime.

\section{Single-mode model of spin-cat state creation via one-axis twisting} \label{sec:SM}

\subsection{Deriving the One-Axis Twisting Hamiltonian}

\begin{figure*}
\centering
\includegraphics[width=1\textwidth]{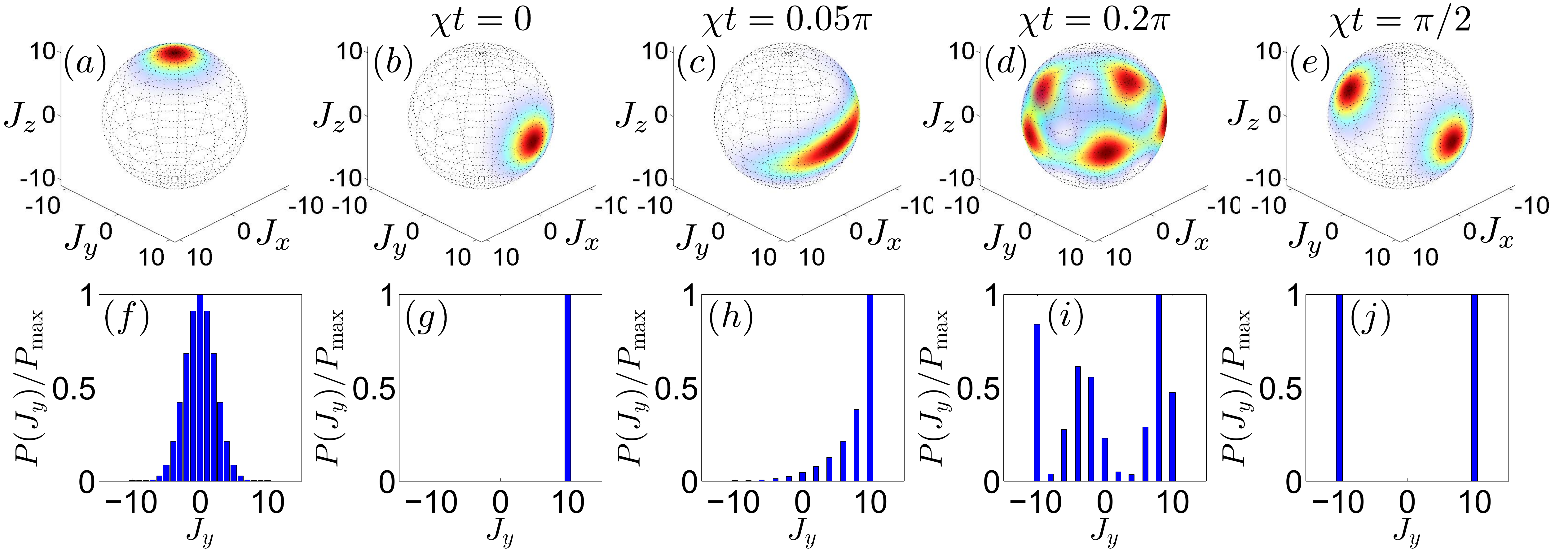}
\caption{\textbf{(a-e)} Single-mode $Q$-functions of $N=20$ atoms ($Q/Q_{\mathrm{max}}$ with $Q(\theta, \phi) = \abs{\langle \alpha(\theta, \phi)|\Psi(t) \rangle}$), showing evolution under one-axis twisting for the state $|\Psi(t)\rangle = \exp(-i \Jhat_z^2 \chi t)|\alpha(\pi/2, \pi/2) \rangle$. (\textbf{f-j}): the corresponding probability distributions in the $\hat{J}_y$ eigenbasis. The system is prepared entirely in a single component ($|\Psi\rangle = |\alpha(0,0) \rangle \equiv |N/2\rangle$) \textbf{(a,f)} , before a $\pi/2$ pulse places the state on the equator of the Bloch sphere ($|\Psi\rangle = e^{-i\Jhat_x\pi,2}|N/2\rangle = |\alpha(\pi/2,\pi/2) \rangle$) \textbf{(b,g)}. Using this as the initial state, the one-axis twisting interaction [Eq.~\eqref{eq:Hoat}] creates a nonclassical state \textbf{(c,h)}, and quickly reaches the over-squeezed regime \textbf{(d,i)}. Eventually, after $t_{\mathrm{cat}}=\pi/2\chi$ the state becomes a spin-cat state in the $\hat{J}_y$ basis \textbf{(e,j)}.}
\label{fig:scheme}
\end{figure*}

The physical system we consider is a two-component Bose-Einstein condensate (BEC), with components labelled $a$ and $b$. In terms of the bosonic field operators $\hat{\psi}_j(\mathbf{r})$, which obey commutation relations $[\hat{\psi}_j(\mathbf{r}), \hat{\psi}^\dagger_k(\mathbf{r}')]=\delta(\mathbf{r}-\mathbf{r}') \delta_{jk}$, the full multi-mode Hamiltonian for the system is 
\begin{eqnarray} \label{eq:chap7Ham} 
\hat{\mathcal{H}} &=& \sum_{j=a,b} \int d \mathbf{r} \hat{\psi}_j^\dagger(\mathbf{r}) H_0 \hat{\psi}_j(\mathbf{r}) \nonumber  \\ 
&+& \sum_{j,k=a,b} \frac{g_{jk}}{2} \int d \mathbf{r} \hat{\psi}_j^\dagger(\mathbf{r}) \hat{\psi}_k^\dagger(\mathbf{r}) \hat{\psi}_j(\mathbf{r}) \hat{\psi}_k(\mathbf{r}) , 
\end{eqnarray}
where $H_0 = \hat{p}^2/2M + V(\mathbf{r})$ is the single-particle Hamiltonian (momentum operator $\hat{p}$, mass $M$, and external trapping potential $V(\mathbf{r})$) and $g_{jk}=4\pi \hbar^2 a_{jk}/M$ is the interaction strength for $s$-wave scattering length $a_{jk}$.

It is common to study this system within the single-mode approximation, which assumes each mode is well described by the same wavefunction $\phi_j(\mathbf{r})$, with $j=a,b$. This is a good approximation for sufficiently small, tightly trapped condensates that the motional dynamics are effectively ``frozen'' over timescales of interest, and may be integrated out \cite{Esteve2008}. Quantitatively, the single-mode Hamiltonian is obtained by making the approximation

\begin{eqnarray}
\hat{\psi}_a(\mathbf{r}) &\approx& \hat{a} \phi_a(\mathbf{r}) \label{sm_ansatz1} \\
\hat{\psi}_b(\mathbf{r}) &\approx& \hat{b} \phi_b(\mathbf{r}) \, , \label{sm_ansatz2}
\end{eqnarray}
with $\left[\ahat, \ahatd\right] = \left[\bhat, \bhatd\right] =1$, $\left[\ahat, \bhat \right] = \left[\ahat, \bhatd\right] =\left[\ahatd, \bhat\right]=0$. 
The wavefunctions $\phi_j(\mathbf{r})$ are normalised to unity. 

A convenient description for a system of $N$ conserved, two-level bosons is the SU(2) angular momentum algebra. In terms of ladder operators $\hat{J}_+=\hat{a} \hat{b}^\dagger$, $\hat{J}_-=(\hat{J}_+)^\dagger$ and number operators $\hat{N}_a=\hat{a}^\dagger \hat{a}$, $\hat{N}_b=\hat{b}^\dagger \hat{b}$, the population difference $\hat{J}_z=(\hat{N}_a-\hat{N}_b)/2$ obeys the standard angular momentum commutation relations with $\hat{J}_x=(\hat{J}_++\hat{J}_-)/2$ and $\hat{J}_y=-i(\hat{J}_+-\hat{J}_-)/2$.

Neglecting the single-particle energies, which amount only to a trivial rotation, in the single-mode regime the multi-mode Hamiltonian [Eq.~\eqref{eq:chap7Ham}] becomes
\begin{equation}
\mathcal{H} = \hbar \chi_{aa} \hat{N}_a^2 + \hbar \chi_{bb} \hat{N}_b^2 + 2 \hbar \chi_{ab} \hat{N}_a \hat{N}_b + \hbar \chi_{aa} \hat{N}_a + \hbar \chi_{bb} \hat{N}_a  \, ,
\end{equation}
where
\begin{equation} \label{eq:chiint}
\chi_{jk}=\frac{g_{jk}}{2} \int d\mathbf{r} |\phi_j(\mathbf{r})|^2 |\phi_k(\mathbf{r})|^2 \, , 
\end{equation}
with $j,k=a,b$. To account for asymmetric interactions $\chi_{aa} \neq \chi_{bb}$, we insert a $\pi$-pulse half-way through the evolution, which is described by the unitary  operator $\hat{U}_{\pi}=\exp({-i \hat{J}_x \pi})$. The single-mode time evolution is generated by
\begin{align}  
\hat{U}(t) &= e^{-i \hat{H} t/2} \hat{U}_{\pi} e^{-i \hat{H} t/2} ,
\end{align}
which can be re-written
\begin{equation}
\hat{U}(t)=\hat{U}_\pi \hat{U}_N e^{-i \hat{J}_z^2 \chi t}  ,
\end{equation}
with the effective twisting rate 
\begin{equation} \label{eq:chidef}
\chi = \chi_{aa}+\chi_{bb}-2\chi_{ab} .
\end{equation}
We have also defined the unitary
\begin{equation}
\hat{U}_N=e^{-i t \left[ \frac{1}{4}(\chi_{aa}+\chi_{bb}+2\chi_{ab})\hat{N}^2+\frac{1}{2} (\chi_{aa}+\chi_{bb}) \hat{N} \right]} ,
\end{equation}
which in an SU(2) system amounts only to a global phase, and is henceforth neglected. The $\pi$-pulse is similarly unimportant as it only reverses the sign of $\hat{J}_z$ and $\hat{J}_y$. Thus, the single-mode Hamiltonian is equivalent to the well known one-axis twisting (OAT) interaction,
\begin{equation} \label{eq:Hoat}
\hat{H}= \hbar \chi \hat{J}_z^2 \equiv \hat{H}_{\mathrm{OAT}} .
\end{equation}

\subsection{Creating Spin-Cat States with One-Axis Twisting}
The OAT Hamiltonian is capable of producing macroscopic superpositions of collective-spin eigenstates. It is convenient to work in the $\hat{J}_z$ eigenbasis, $\hat{J}_z |m \rangle = m |m \rangle$ for $J=N/2$, where $N$ is  total number of atoms and $m$ is half the population difference. We begin with a separable, coherent spin state \cite{Radcliffe1971, Gross2012} $\ket{\alpha(\theta, \phi)}$, defined by
\begin{eqnarray}
\ket{\alpha(\theta, \phi)} &=& e^{i\phi\Jhat_z}e^{i\theta \Jhat_y} |N/2\rangle \nonumber \\
&=& \sum_{m=-J}^J C^J_m(\theta)  e^{-i (J+m)\phi} |m \rangle \, ,
\end{eqnarray}
where 
\begin{equation}
C^J_m(\theta) = \begin{pmatrix} 2J \\ J+m \end{pmatrix}^{1/2} \cos(\theta/2)^{J-m} \sin(\theta/2)^{J+m}.
\end{equation}
Evolving this state under $\hat{H}_{\mathrm{OAT}}$ results in a non-linear rotation of each $\hat{J}_z$ component about the $\hat{J}_z$ axis by twisting angle $\chi t$, as illustrated in Fig.~\ref{fig:scheme}(b-e). For small interaction times $\chi t$, the resultant nonclassical state has significantly modified noise properties, and leads to spin squeezing, Fig. \ref{fig:scheme}c). At larger times, the state becomes non-Gaussian, as illustrated in Fig.~\ref{fig:scheme}d). At time
\begin{equation} \label{eq:tcat} 
t_{\mathrm{cat}}=\frac{\pi}{2 \chi} \, ,
\end{equation}
noting that for any integers $J$ and $m$ \footnote{For odd values of $N$ (and therefore half-integer $J$), the final expression is \unexpanded{$e^{-i\hat{J}_z^2\pi/2}|\alpha(\theta, \phi)\rangle =  \frac{e^{-i\pi/4}}{\sqrt{2}}(|\alpha(\theta, \phi+\pi/2)\rangle + i(-1)^{J+\frac{1}{2}} |\alpha(\theta, \phi-\pi/2)\rangle )$}},
\begin{equation}
e^{-i m^2 \pi/2} = \frac{e^{-i\pi/4}}{\sqrt{2}}(1 + i(-1)^J e^{ i(J+m)\pi})\, 
\end{equation}
such that
\begin{widetext}
\begin{eqnarray}
e^{-i \hat{J}_z^2 \pi/2}\ket{ \alpha(\theta, \phi) }  &=&\frac{e^{-i\pi/4}}{\sqrt{2}} \sum_{m=-J}^J C^J_m(\theta) (1 + i(-1)^J e^{i(J+m)\pi}) e^{-i (J+m)\phi} |m \rangle \nonumber \\
&=& \frac{e^{-i\pi/4}}{\sqrt{2}} \Big( \sum_{m=-J}^J C^J_m(\theta) e^{- i(J+m)\phi}\ket{m} + i(-1)^J \sum_{m=-J}^J C^J_m(\theta) e^{- i(J+m)(\phi+\pi)}\ket{m} \Big) \nonumber \\
&=& \frac{e^{-i\pi/4}}{\sqrt{2}}\left(\ket{\alpha(\theta, \phi)} + i(-1)^J\ket{\alpha(\theta, \phi+\pi)}  \right) \, .
\end{eqnarray}
\end{widetext}

\begin{figure}
\centering
\includegraphics[width=\columnwidth]{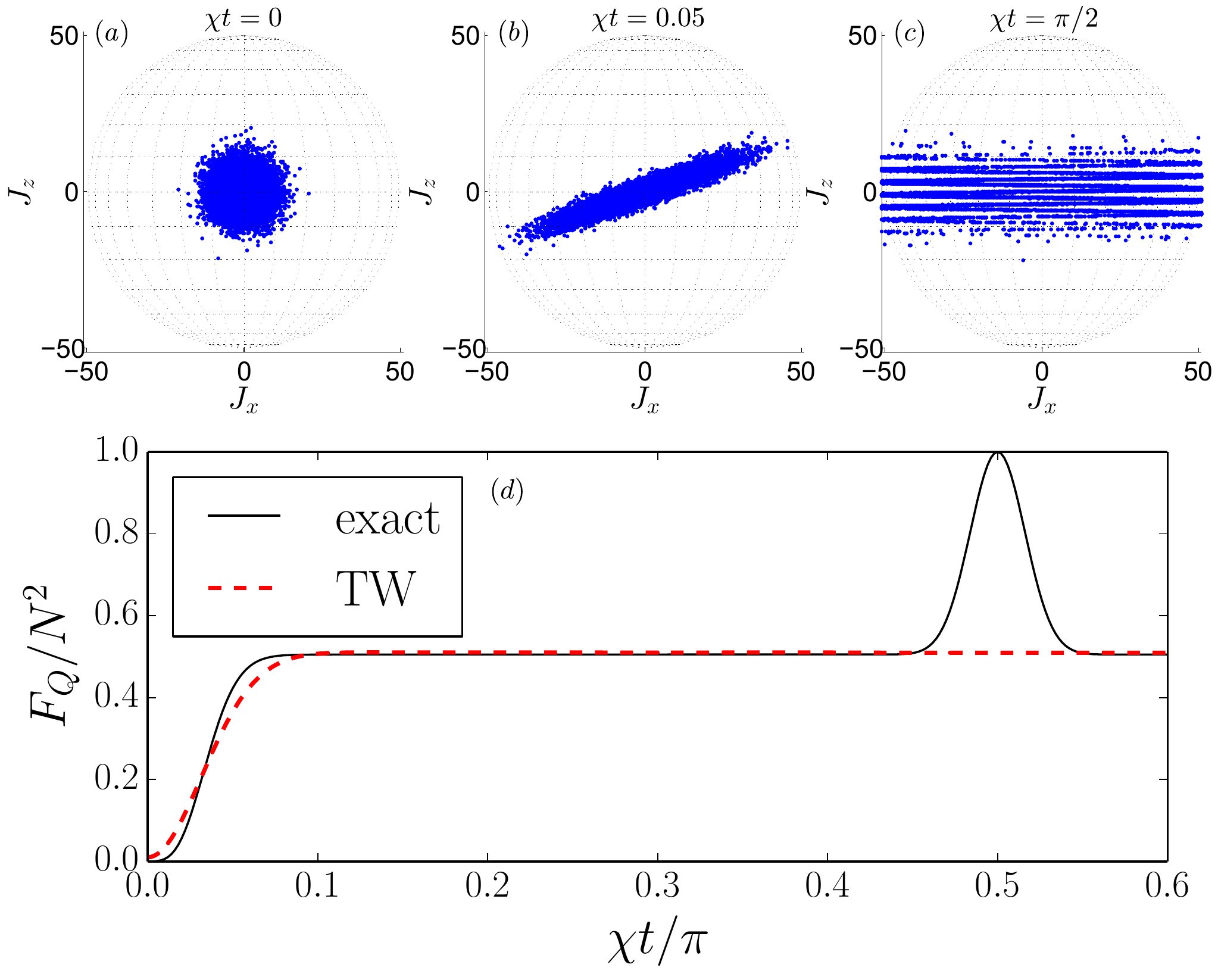}
\caption{QFI vs. time for the state $|\psi(t)\rangle = e^{-i \Jhat_z^2 \chi t }|\alpha(\pi/2,0)\rangle$, calculated analytically (Ref. \cite{Kitagawa1993}, solid black), and with the widely used truncated Wigner method (TW, dashed red). At $\chi t=\pi/2$ the QFI exhibits a ``cat peak'' in the QFI, associated with the creation of a spin-cat state, which is absent in the TW calculation.} 
\label{fig:QFI_sm}
\end{figure}

When we choose $\theta = \pi/2$ (as in Fig.~\ref{fig:scheme}), the state is an equal superposition of the maximal and minimal eigenstates of $\Jhat_y$ (Fig.~\ref{fig:scheme}e). This state is characterised by its large quantum Fisher information (QFI) with respect to the pseudo-spin operator $\hat{J}_y$: $F_Q = 4 \mathrm{Var}(J_y) = N^2$. That is, for pure state $|\psi\rangle$ under evolution $|\psi_\Omega\rangle = \exp(i\Jhat_y \Omega)|\psi\rangle$, the parameter $\Omega$ may be estimated with Heisenberg-limited sensitivity $\Delta \Omega = 1/\sqrt{F_Q} = 1/N$ \cite{Paris2009, Toth2014, DemkowiczDobrzanski2015}. If the scattering lengths of the two components are asymmetric, evolution under the full Hamiltonian Eq.~\eqref{eq:chap7Ham} may result in drift in the $J_x$, $J_y$ plane. Rather than manually accounting for this drift, it is simpler to calculate the QFI by finding the maximum eigenvalue of the collective-covariance matrix, 
\begin{equation}
F_{i,j} =  2\langle \hat{J}_i \hat{J}_j + \hat{J}_j \hat{J}_i \rangle - 4 \langle \hat{J}_i \rangle \langle \hat{J}_j \rangle . \label{Fij}
\end{equation}
where $i,j=x,y,z$ \cite{Hyllus2010}. The QFI of the state $|\psi(t)\rangle = e^{-i\chi t \Jhat_z^2}|\alpha(\pi/2,0)\rangle$ as a function of time is shown in Fig.~\ref{fig:QFI_sm}. The QFI initially increases rapidly, before quickly reaching a plateau at $F_Q = N^2/2$. At $\chi t = \pi/2$, the state briefly revives to a cat state, and the QFI peaks at $F_Q = N^2$. Throughout this paper, we refer to this peak as the ``cat peak", and take it as the signature of a cat-like state.  

\section{One-dimensional multi-mode model}\label{sec:castinsinatra}

In cases with small atom-number and very tightly confined potentials, such as \cite{Gross2010}, the motional dynamics of the condensate are negligible and the single-mode approximation [\eq{sm_ansatz1} and \eq{sm_ansatz2}] is sufficient to model the evolution of quantum correlations in the system. However, there are cases when the multimode dynamics cause a break-down of the single mode approximation, and we must model the system in a way that accounts for both the quantum correlations and multimode dynamics. The truncated Wigner (TW) method \cite{Drummond1993, Werner1995, Steel1998, Sinatra2000, Sinatra2001, Sinatra2002, Gardiner2002} has been used successfully to model spin-squeezing in the presence of significant multimode dynamics \cite{Haine2009, Sinatra2011, Haine2011, Opanchuk:2012, Haine2014, Nolan2016, Haine:2018, Laudat:2018}. However, the truncation of third order terms in the Focker-Planck equation limits the dynamics to states with positive Wigner functions \cite{gardiner_zoller_book_04}, and therefore cannot be used to model the creation of spin-cat states, which display significant negativity \cite{Pezze:2016_review}. Fig.~\ref{fig:QFI_sm} compares a single-mode TW simulation to the exact calculation. While TW simulation agrees quite well for times less than $\chi t \sim \pi/2$, it does not describe the revival of the state and the associated cat peak. Other phase-space methods also fail, such as positive-P \cite{Drummond1980, Gardiner1993, Deuar2006}, which is restricted to evolution times much less than $t_{\mathrm{cat}}$ due to the exponential divergence of stochastic trajectories, or number-phase Wigner \cite{Hush2010, Hush2012}, which is negative for coherent-spin states.  We circumvent these issues by employing the method of Sinatra and Castin \cite{Sinatra2000, Li2009, Kurkjian2017}, which is described in depth in Appendix \ref{sec:appA}. Briefly, the idea is to expand the state in the number basis, and then evolve the wavefunction for \emph{each} number component within the Hartree-Fock approximation. In this way it is possible to capture both multi-mode dynamics and quantum correlations. We define a new Fock space with bosonic annihilation operators
\begin{align} \label{eq:CSladder}
\hat{a}_{\phi_{a,m}} &= \int d \boldr \phi_{a,m}^*(\boldr, t) \hat{\psi}_a(\boldr) \\
\hat{b}_{\phi_{b,m}} &= \int d \boldr \phi_{b,m}^*(\boldr, t) \hat{\psi}_b(\boldr) ,
\end{align}
which are used to construct a set of \emph{dynamic} basis states, labelled by $m$ 
\begin{align} \label{eq:chap7state}
& |m; \phi_{a,m}(\boldr, t), \phi_{b,m}(\boldr, t) \rangle = \\
& e^{-i A_m(t)/\hbar}\frac{\left(\hat{a}^\dagger_{\phi_{a,m}} \right)^{n_a} }{\sqrt{n_a!}}\frac{\left(\hat{b}^\dagger_{\phi_{b,m}} \right)^{n_b}}{\sqrt{n_b!}} |0\rangle \nonumber , 
\end{align}
with $n_a = N/2 + m$, $n_b = N/2-m$, and we implicitly assume a fixed total number $n_a + n_b = N$. These basis states are SU(2) states with respect to the multi-mode pseudo-spin operators $\Jhat_x = (\Jhat_+ + \Jhat_-)/2$, $\Jhat_y = -i(\Jhat_+ - \Jhat_-)/2$, $\hat{J}_z=(\hat{N}_a-\hat{N}_b)/2$ with
\begin{align}
\hat{J}_+ &= \int d \xi \hat{\psi}_a(\xi) \hat{\psi}_b^\dagger(\xi) \\
\hat{N}_j &= \int d \xi \hat{\psi}_j^\dagger(\xi) \hat{\psi}_j(\xi) ,
\end{align}
and $\hat{J}_-=(\hat{J}_+)^\dagger$. The approximation in this method is that each number-component is well described by a single wavefunction. Although this ansatz assumes a Hartree product state for each number component, it is able to capture quantum correlations \emph{between} number components that may arise from subsequent dynamics. 

\begin{figure*} 
\centering
\includegraphics[width=\textwidth]{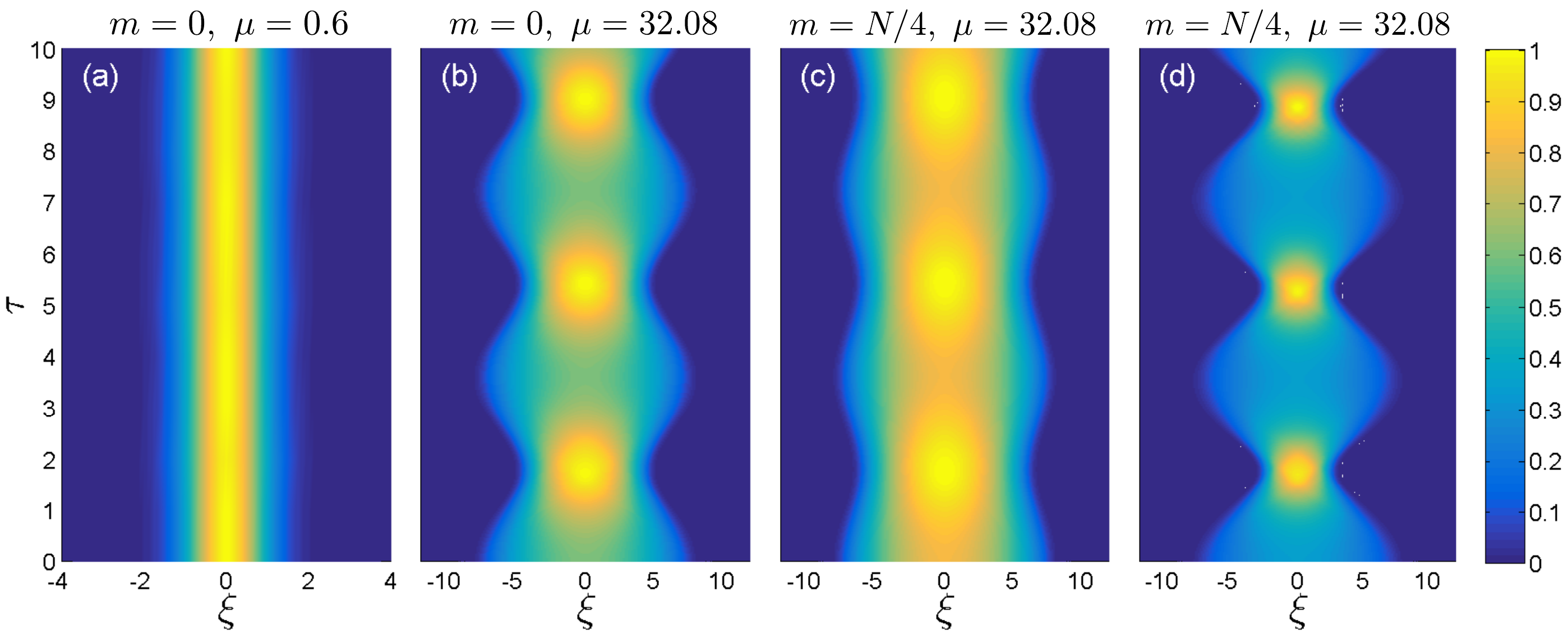}
\caption{Dynamics of $|\phi_{a,m}|^2/|\phi_{\mathrm{max}}|^2$ \textbf{(a-c)} and $|\phi_{b,m}|^2/|\phi_{\mathrm{max}}|^2$ \textbf{(d)}, labelled by $m=(n_a-n_b)/2$ with, $N=100$, $\kappa=0$ and $\lambda=1$. \textbf{(a)} When the chemical potential is close to the single-particle ground state energy the system is well described by a single-mode treatment, i.e. the dynamics are almost identical for populations $a$,$b$ and for all number components $m$. \textbf{(b)} A larger chemical potential gives rise to breathing dynamics shown here for the $m=0$ number component, which is different for each $m$, shown in \textbf{(c)}. Multi-mode dynamics also differ between the components, for instance \textbf{(d)} shows the density $|\phi_{b,m}|^2/|\phi_{\mathrm{max}}|^2$. For $m = N/4$, $|\phi_{b,m}|^2$ differs significantly from $|\phi_{a,m}|^2$. The reason is that $N_a = (N/2 + m) = 3/4 N$, while $N_b = (N/2 -m) = N/4$, such that the initial condition for $\phi_{b,m}$ is significantly further from a stationary state of Eq.~\eqref{eq:chap7des1}.}
\label{fig:densities}
\end{figure*}

We consider a harmonically trapped BEC, where the trapping frequency $\omega$ along the $x$ axis is small compared to the $y$ and $z$ directions. The dynamics in the transverse dimensions are therefore integrated out resulting in an effective one dimensional (1D) system, with modified interaction strengths $\tilde{g}_{ij}$. In practice, $\tilde{g}_{ij}$ is determined by the degree of transverse confinement. However, in a 1D model, the dynamics are entirely determined by the magnitude of the chemical potential, $\mu_0$, relative to the non-interacting ground-state energy $\hbar \omega/2$. Thus, the relevant parameter is the dimensionless chemical potential $\mu = \mu_0/\hbar \omega$. As such, we adjust $\tilde{g}_{ij}$ such that the chemical potential matches the chemical potential of the 3D system we wish to emulate. Working in dimensionless oscillator units $\tau = \omega t$ and $\xi=x\sqrt{M \omega/\hbar}$, assuming a Hamiltonian of the form \eq{eq:chap7Ham}, the (unity-normalised) mode-function $\phi_{j,m}(\xi, \tau)$ and phase factors $A_m(\tau)$ that define the states Eq.~\eqref{eq:chap7state} evolve in time under the equations (with $j=a,b$)
\begin{align}  \label{eq:chap7des1}
i \frac{\partial  \phi_{j,m}}{\partial \tau} &= \left(-\frac{1}{2} \frac{\partial^2}{\partial \xi^2} + \frac{1}{2} \xi^2\right) \phi_{j,m} \notag \\
&+ \left(\tilde{g}_{jj} (n_j-1)|\phi_{j,m}|^2 + \tilde{g}_{jk} n_k |\phi_{k,m}|^2 \right) \phi_{j,m} \\
\nonumber \\ \label{eq:chap7des2}
\frac{dA_m}{d\tau} &= -\sum_{j=a,b} \frac{\tilde{g}_{jj}}{2}n_j(n_j-1) \int d\xi |\phi_{j,m}|^4 \notag \\
& - \tilde{g}_{ab} n_a n_b \int d\xi |\phi_{a,m}|^2 |\phi_{b,m}|^2 
\end{align}
(see Appendix \ref{sec:appA} for derivation). If all the atoms are in mode $a$, before a $\pi/2$ rotation about the $J_x$ axis instantaneously puts each atom in an equal superposition of state $a$ and $b$ (equivalent to the maximal $J_y$ eigenstate), then the appropriate initial conditions are 
\begin{align}
\phi_{a,m}(\xi, 0)&=\phi_{b,m}(\xi, 0)=\phi_0(\xi), \\
A_m(0) &= 0 
\end{align}
where $\phi_0(\xi)$ is the solution to
\begin{equation} \label{eq:chap7ground}
\mu \phi_0= \left( -\frac{1}{2} \frac{\partial^2}{\partial \xi^2} + \frac{1}{2} \xi^2 + N \tilde{g}_{aa} |\phi_0|^2 \right) \phi_0.
\end{equation}
which represents the ground state wavefunction of the system when all $N$ particles are initially in mode $a$ (ie, $m= N$) with chemical potential $\mu$. With these assumptions, the quantum state of the system is given by
\begin{align} \label{eq:initialstate}
&|\psi(\tau) \rangle = \nonumber \\
&\sum_{m=-N/2}^{N/2} \sqrt{\frac{N!}{n_a!n_b!}} (c_a)^{n_a} (c_b)^{n_b}  |m; \phi_{a,m}(\xi, \tau), \phi_{b,m}(\xi, \tau) \rangle \, ,
\end{align}
with $c_a = \frac{1}{\sqrt{2}}$ and $c_b = \frac{i}{\sqrt{2}}$. Throughout this paper we write the interaction strengths in terms of dimensionless parameters,
\begin{align}
\tilde{g}_{aa} &= g_0 \\ 
\tilde{g}_{bb} &= \lambda g_0\\
\tilde{g}_{ab} &= \kappa g_0 ,
\end{align}
where $g_0$ is chosen to determine a particular value of $\mu$, for some $N$. This is done by solving Eq.~\eqref{eq:chap7ground}, and imposing $\int d \xi |\phi_0 (\xi)|^2=1$. 

Fig.~\ref{fig:densities} shows the magnitude of the mode-functions $|\phi_{j,m}(\xi,\tau)|^2$ as a function of time. We chose  $N=100$, for several values of $\mu$. For $\mu=0.6$, which is just slightly above the non-interacting ground state energy $E_0 = \frac{1}{2}$, the density is approximately static. Furthermore, it is the same for all $m$ components, which is the essence of the single-mode approximation. However, when we increase the chemical potential to $\mu = 32.08$ (with $\lambda=1$ and $\kappa = 0$), we observed significant breathing dynamics in the mode-function shape. Even in the case of symmetric interactions ($\lambda=1$), when $m \neq 0$, $\phi_{a,m}(\xi)$ and $\phi_{b,m}(\xi)$ breathe with different amplitudes, which will reduce the spatial overlap of these components. In fact when $\lambda=1$,  $\phi_{a,m}(\xi,\tau)=\phi_{b,-m}(\xi, \tau)$.

\section{Effect of multimode dynamics on the QFI}\label{sec:MMQFI}

\begin{figure} [h!]
\centering
\includegraphics[width=\columnwidth]{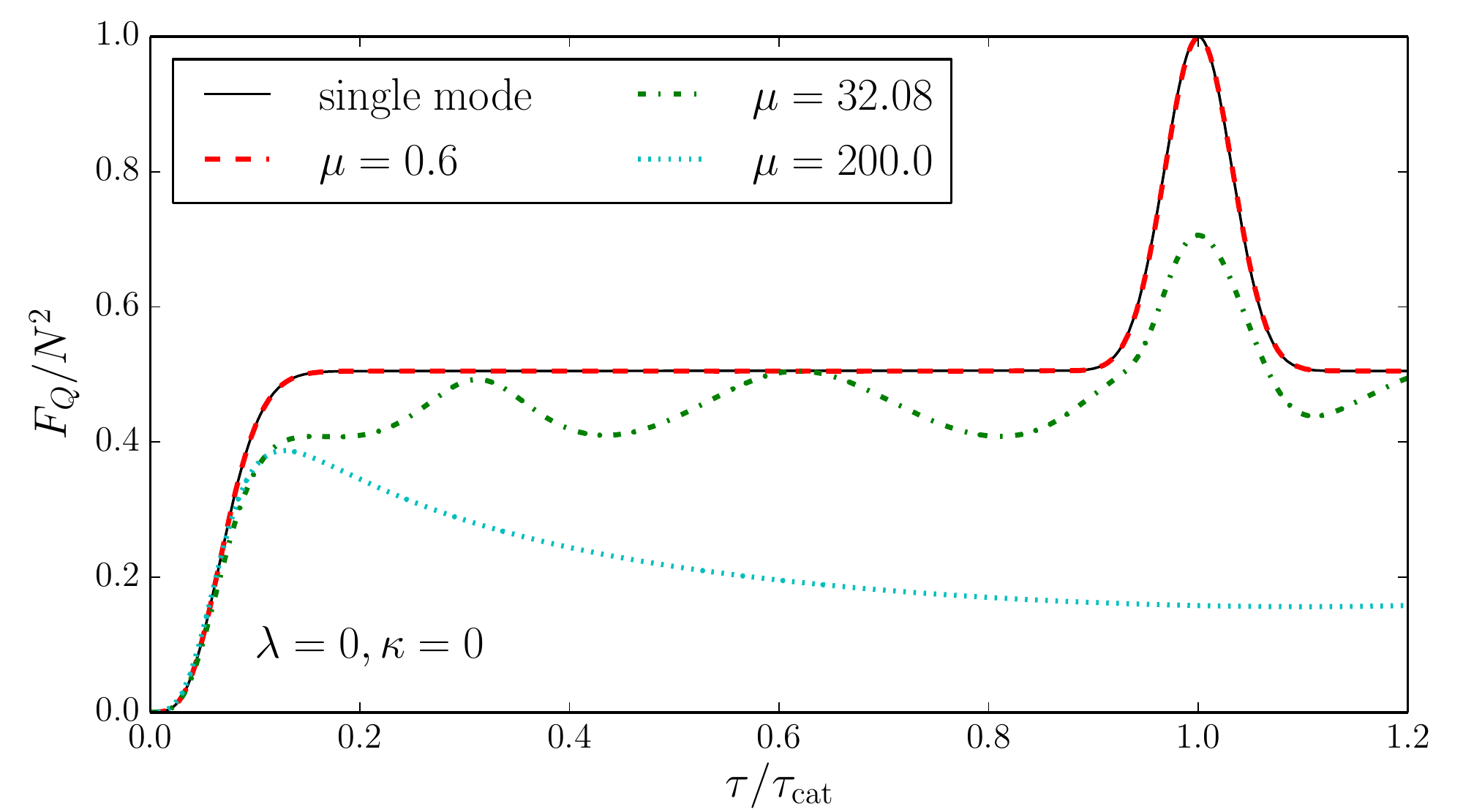}
\caption{The quantum Fisher information of a maximally non-linear ($\kappa=0$), symmetric ($\lambda=1$) condensate with $N=100$ as a function of evolution time for several $\mu$. Times $\tau_{\mathrm{cat}}$ were taken directly from numerics by finding the peak, except for $\mu=200.0$ which was estimated from Eq.~\eqref{eq:chiTF} and Eq.~\eqref{eq:tcat}. The times are $\tau_{\mathrm{cat}}=1570$, $5.735$, $0.9817$ for $\mu=0.6$, $32.08$, $200.0$, respectively. Chemical potential $\mu$ is in units of $\hbar \omega$. }
\label{fig:catpeaks}
\end{figure}

Figure \ref{fig:catpeaks} shows the QFI, calculated via \eq{Fij}. When $\mu =0.6 $, we see excellent agreement with the single mode model. However, despite a clearly defined cat peak with maximum amplitude $F_Q=N^2$ the time taken for this revival is large; $t_{\mathrm{cat}} \approx 1570/\omega$ for $\mu=0.6$. Increasing $\mu$ significantly decreases $t_{\mathrm{cat}}$, however the peak QFI is diminished, eventually disappearing completely.

We can understand the scaling of $t_\mathrm{cat}$ with $\mu$ by considering the shape of $|\phi_0(\xi)|^2$. In the regime where $\mu \gg \hbar \omega$, the ground state is well approximated by the Thomas-Fermi (TF) solution \cite{Dalfovo1999}
\begin{equation}
|\phi_0(\xi)|^2 = \frac{\mu-\frac{1}{2}\xi^2}{Ng_0} 
\end{equation}
for $\xi^2 < 2\mu$, and $0$ otherwise, and 
\begin{equation}
g_0 = \frac{4\sqrt{2}\mu^{\frac{3}{2}}}{3N}
\end{equation}
is chosen to enforce the normalisation of $|\phi_0(\xi)|^2$. This is equivalent to choosing a particular value of $g_0$ to determine $\mu$. 

\begin{figure}
\centering
\includegraphics[width=\columnwidth]{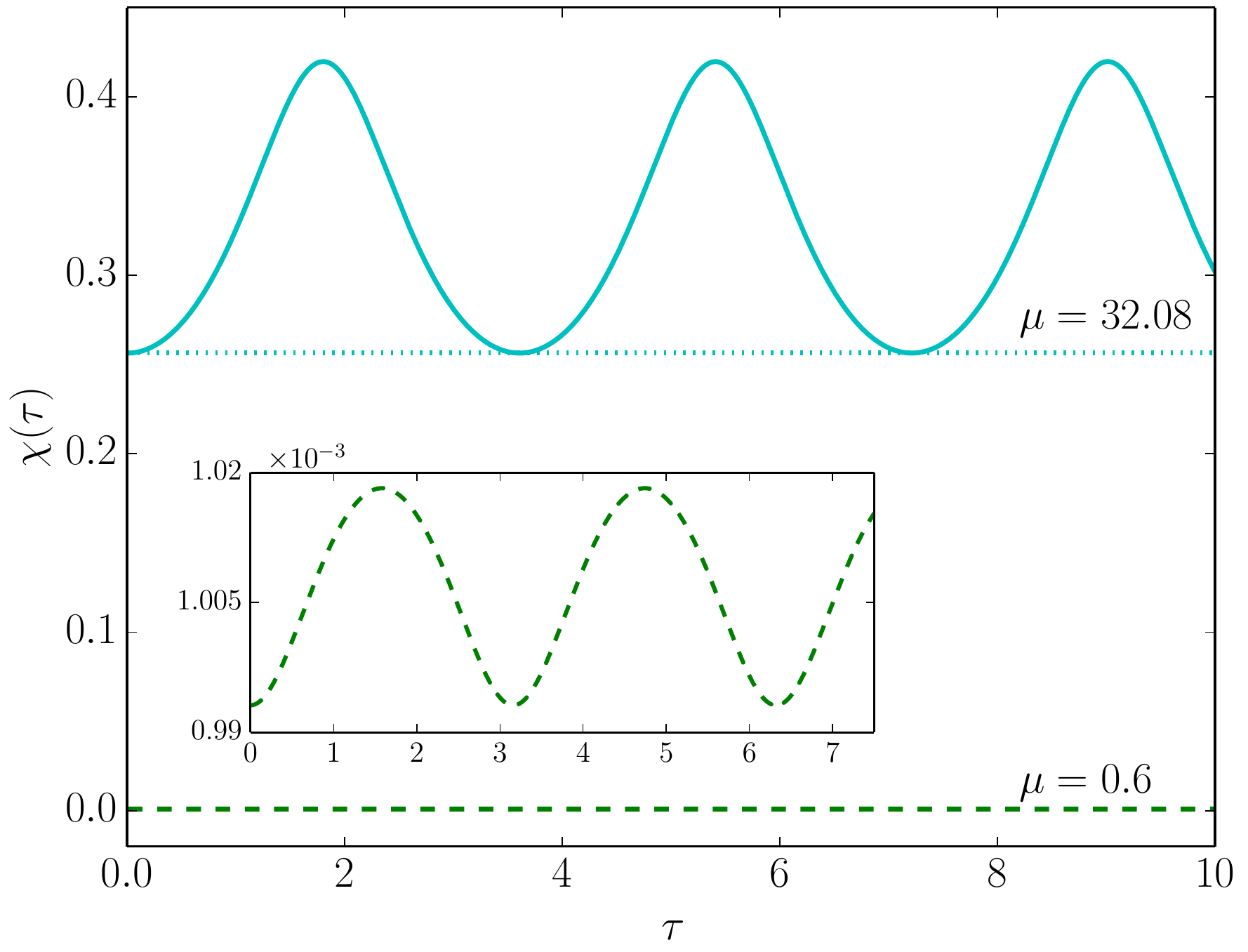}
\caption{The 1D twisting rate $\chi$ under breathing dynamics with $\lambda=1$ and $\kappa=0$, far from the single-mode regime $\mu=32.08$ (solid cyan) and close to the single-mode regime $\mu=0.6$ (dashed green), which is close to 0. We also include twisting rate $\chi$ of the TF ground state [Eq.~\eqref{eq:chiTF}] (dashed cyan), which agrees excellently with the numeric result at $\tau=0$. Inset: Magnification of $\mu=0.6$ data which shows small, but non-zero dynamics. In the TF regime the period of the breathing motion agrees well with the $n=2$ excitation period, Eq.~\eqref{eq:Tn}.}
\label{fig:chi}
\end{figure}

Inserting this into \eq{eq:chiint} and \eq{eq:chidef}, with $\phi_a(\xi) = \phi_b(\xi) = \phi_0(\xi)$ gives the twisting rate of the TF initial state
\begin{equation} \label{eq:chiTF}
\chi_\mathrm{TF} = (1 + \lambda - 2\kappa) \frac{2\mu}{5N} \, ,
\end{equation}
and therefore
\begin{equation} \label{eq:tcatTF}
\tau_\mathrm{cat} =\omega t_\mathrm{cat} = \frac{5\pi N}{4\mu(1+\lambda-2\kappa)} .
\end{equation}
Subsequent dynamics in the twisting rate $\chi(\tau)$ occur when $\lambda, \kappa \neq 1$. For symmetric scattering lengths ($\lambda=1$) the excitation spectrum of the two component TF ground-state agrees well with the 1D single component result presented in the appendix of Ref. \cite{Kneer1998}, which finds oscillating excitations with period
\begin{equation} \label{eq:Tn}
T_{n} = \frac{2 \pi}{\sqrt{\frac{n}{2}(n+1)}} .
\end{equation} 
Breathing motion is the $n=2$ excitation, ($T_2 \approx 3.63$) which agrees well with the dynamics observed in Figure \ref{fig:densities}. Figure \ref{fig:chi} compares $\chi_{\mathrm{TF}}$ to $\chi(\tau)$ calculated via \eq{eq:chiint} using the mean-field wavefunctions, obtained by setting $m=0$ in \eq{eq:chap7des1}, i.e $\phi_j(\xi, \tau) = \phi_{j,0}(\xi, \tau)$. In the TF regime we observe breathing oscillations in $\chi(\tau)$ with period $T_2 \approx 3.63$, and find good agreement between $\chi(\tau)$ and $\chi_{\mathrm{TF}}$ at $\tau = 0$. As $\mu$ approaches the single-particle energy $E_0 = \frac{1}{2}$, the multi-mode dynamics vanish, but so does the effective twisting rate $\chi(\tau)$. 

\begin{figure*}
\centering
\includegraphics[width=\textwidth]{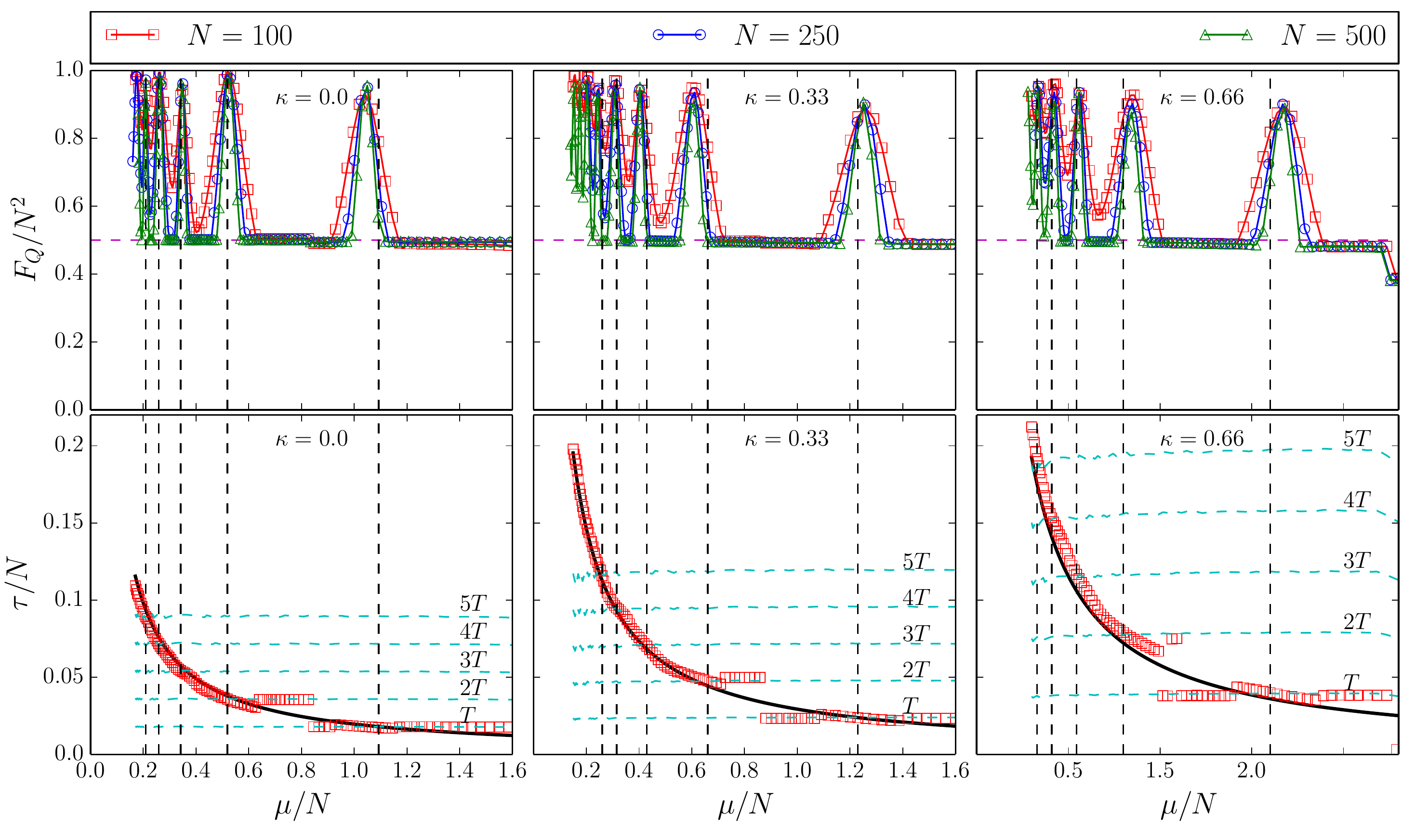}
\caption{\textbf{(top row)} For a symmetric condensate $\lambda=1$, the maximum QFI is displayed for a range of parameters. The QFI is compared to $F_Q=N^2/2$ (dashed magenta), which is quickly (relative to $\tau_{\mathrm{cat}}$) reached under OAT. \textbf{(bottom row)} For $N=100$, we display the time at which the maximum QFI occurs (not necessarily $t_\mathrm{cat}$). This is compared to the TF result (Eq.~\eqref{eq:tcatTF}, solid black), and deviations from this curve indicate that the corresponding maximum QFI is not associated with a cat peak. The dashed cyan lines are integer multiples of the period of $|\gamma_2^{aa}(m=0)|$. }
\label{fig:fqvsmu}
\end{figure*}

Although Eq.~\eqref{eq:tcatTF} neglects dynamics (Figure \ref{fig:fqvsmu} explores deviations from this formula that arise due to multi-mode dynamics), it is suggestive that $t_{\mathrm{cat}}$ could be reduced by increasing $\mu$. However, in a 1D simulation the time scale (in SI units) generally depends on $\mu$. Despite this, it is possible to meaningfully study the dependence of $t_{\mathrm{cat}}$ on $\mu$ in absolute terms. The 1D and 3D interaction strengths are related by some area $A_{\perp}$, $g_{\mathrm{1D}} = g_{\mathrm{3D}}/A_{\perp}$, and so one could vary $\mu$ by adjusting $A_{\perp}$ but keeping $\omega$ fixed, which would fix the time-scale between simulations with different $\mu$. In this case, so long as the TF approximation holds, $\mu_{\mathrm{TF}} \propto g_{\mathrm{1D}}^{2/3}$, and $t_{\mathrm{cat}}$ would be reduced by relaxing $A_{\perp}$, which supports multi-mode dynamics. 

Figure \ref{fig:catpeaks} reveals a trade-off between this speed-up and the maximum QFI the state reaches, as clearly the multi-mode dynamics have a deleterious effect on the QFI, especially around $\tau_{\mathrm{cat}}$. To explore the cause of this, consider $F_Q=4 \mathrm{Var}(\hat{J}_y)$, which is the optimal generator at $t_\mathrm{cat}$ for symmetric interactions ($\lambda = 1$). It is convenient to define the overlap between the mode functions of different number components 
\begin{equation}
\gamma^{jk}_{m^\prime}(m, \tau) = \int d\xi \phi_{j,m}(\xi, \tau) \phi_{k, m-m^\prime}^*(\xi, \tau) ,
\end{equation}
and to decompose the QFI into terms that depend on the $0$th, $1$st and $2$nd order overlaps respectively \cite{Haine2015a}
\begin{equation}
F=4\mathrm{Var}(\hat{J}_y) = F_0+F_1+F_2 \, ,
\end{equation}
where
\begin{align} \label{eq:F0}
F_0 &= \langle \hat{J}_+ \hat{J}_- \rangle + \langle \hat{J}_- \hat{J}_+ \rangle \\
F_1 &=   - 4 \langle  \hat{J}_y \rangle^2 \\ \label{eq:F2}
F_2 &= - \langle \hat{J}_+ \hat{J}_+ \rangle - \langle \hat{J}_- \hat{J}_- \rangle .
\end{align}
Suppressing the $\tau$ dependence, expressing $F_0$, $F_1$ and $F_2$ in terms of  \eq{eq:initialstate} gives

\begin{widetext}
\begin{subequations}
\begin{align}
F_0 &= N+2 \sum_{n_a=1}^{N-1} \frac{N!}{(n_a-1)! (n_b-1)!} |c_a|^{2n_a} |c_b|^{2n_b} \left| \gamma^{ab}_0(m) \right|^2 \\
F_1 &= - \mathrm{Im} \Bigg( \sum_{n_a=1}^{N} \frac{N!}{(n_a-1)! n_b!} |c_a|^{2(n_a-1)} |c_b|^{2n_b} c_b^* c_a  e^{i [A_{m-1} -A_m]/\hbar} \gamma^{ab}_1(m) \left[ \gamma^{aa}_1(m) \right]^{n_a-1} \left[ \gamma^{bb}_1(m) \right]^{n_b} \Bigg)^2 \\
F_2 &= - \sum_{n_a=2}^{N} \frac{N!}{(n_a-2)! n_b!} |c_a|^{2(n_a-2)} |c_b|^{2n_b} (c_b^*)^2 c_a^2 e^{i [A_{m-2}-A_m]/\hbar}  \left[ \gamma^{ab}_2(m) \right]^2 \left[ \gamma^{aa}_2(m) \right]^{n_a-2} \left[ \gamma^{bb}_2(m) \right]^{n_b}  - c.c  . \label{eq:F2CS}
\end{align}
\end{subequations}
\end{widetext}
where $c.c$ denotes the complex conjugate. 
\begin{figure} [h!]
\centering
\includegraphics[width=\columnwidth]{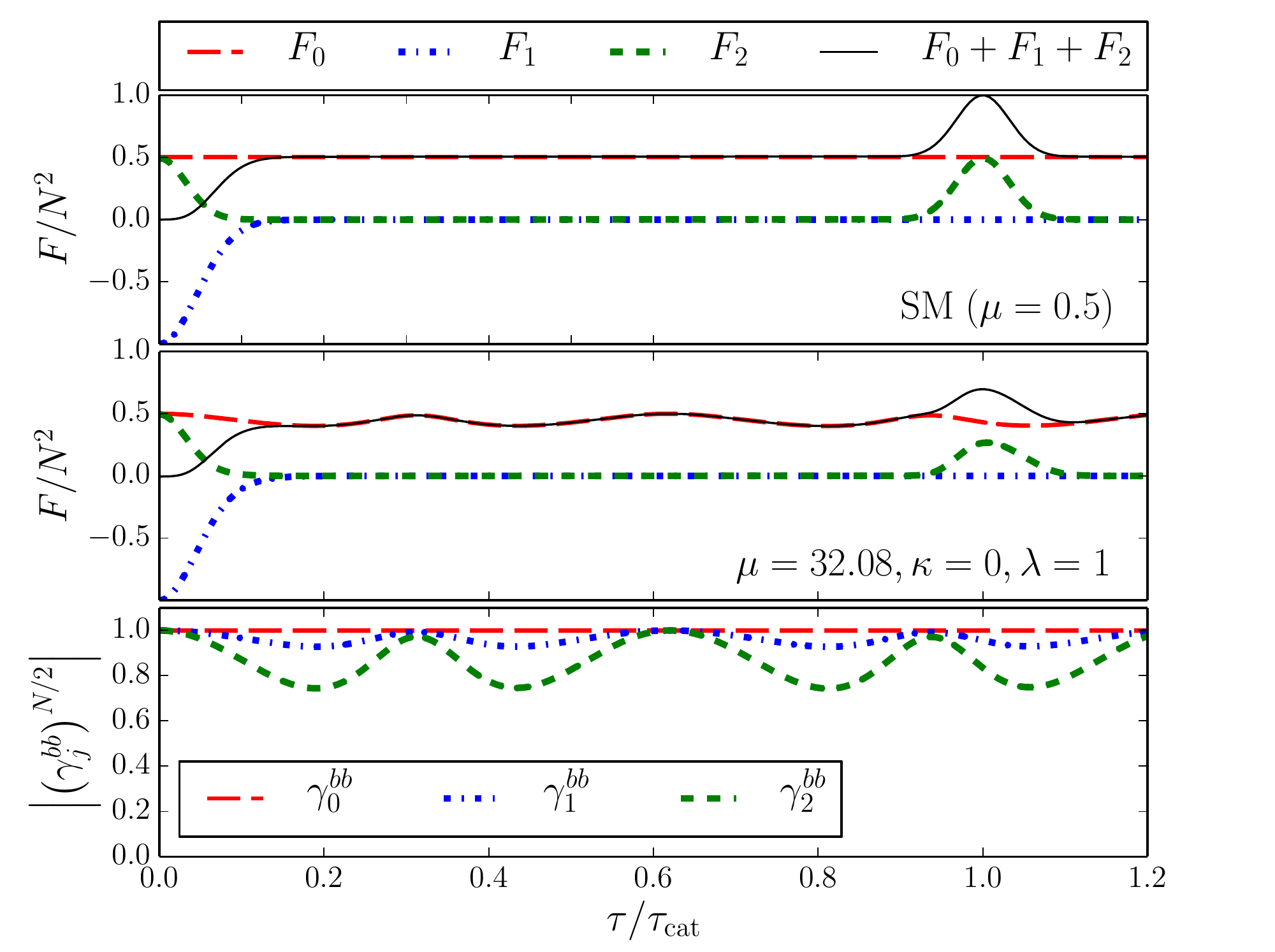}
\caption{Shows $F_0$, $F_1$, $F_2$ and $F_0+F_1+F_2=4\mathrm{Var}(\hat{J}_y)$ for $N=100$ as a function of evolution time for single-mode (SM) dynamics \textbf{(top)} and \textbf{(middle)} a multi-mode simulation performed using the method of Sinatra and Castin. $F_2$ is the term responsible for the cat peak. \textbf{(bottom)} The magnitude of the overlap $[\gamma^{bb}_{j}(m=0, \tau)]^{N/2}$ for $j=1,2,3$, and $\mu=32.08$. Due to the binomial coefficients in Eq.~\eqref{eq:initialstate}, the $m=0$ term contributes the most to $F_2$ [c.f. Eq.~\eqref{eq:F2CS}].}
\label{fig:overlaps}
\end{figure} 

Figure \ref{fig:overlaps} shows these, and their sum, in the single-mode regime (top panel) and in a multi-mode regime (middle panel). In the single-mode regime, $|\gamma_{m^\prime}^{jk}(m,\tau)|^2 = 1$ for all $\tau$. Therefore $F_0$ is conserved and reduces to $F_0=N^2/2+N-2\langle \hat{J}_z^2 \rangle=N^2/2$. In the multi-mode system, $F_0\leq N^2/2$. We also have for any SU(2) system (multi-mode or single-mode) $-N/2 \leq \langle \hat{J}_y \rangle \leq N/2$ and $0 \leq F_0+F_1+F_2=4\mathrm{Var}(\hat{J}_y) \leq N^2$. The first bound implies $-N^2 \leq F_1 \leq 0$, and all of these bounds can be combined to deduce $0 \leq F_2 \leq N^2/2$. The terms $F_1$ and $F_2$ initially decay, and reach $\sim 0$ at approximately $\tau = 0.1 \tau_\mathrm{cat}$. At $\tau = \tau_\mathrm{cat}$, $F_2$ experiences a revival to the maximum value, which is responsible for the cat peak.  The decay of the overlaps is responsible for the decreased QFI in the multimode regime. As such, the QFI at $\tau_\mathrm{cat}$ is independent of $\gamma^{jk}_1(m)$. 

In the multimode regime, the decreased QFI is primarily determined by the decay of $|\gamma_{m^\prime}^{jk}(m,\tau)|^2$.  Figure \ref{fig:overlaps} (bottom row) shows the magnitude of the overlaps to the power of $n_b$ as this directly appears in  Eq.~\eqref{eq:F2CS},  $\gamma^{bb}_0(0)$, $\gamma^{bb}_1(0)$ and $\gamma^{bb}_2(0)$ for the symmetric number component ($m=0$), which has the largest weighting in Eq.~\eqref{eq:initialstate}. The second-order overlap is reduced more than the first or zeroth-order overlaps, and as the first-order overlaps already do not contribute, this indicates that poor second-order overlaps are primarily responsible for the reduction in maximum QFI. 
\begin{figure} [h!]
\centering
\includegraphics[width=\columnwidth]{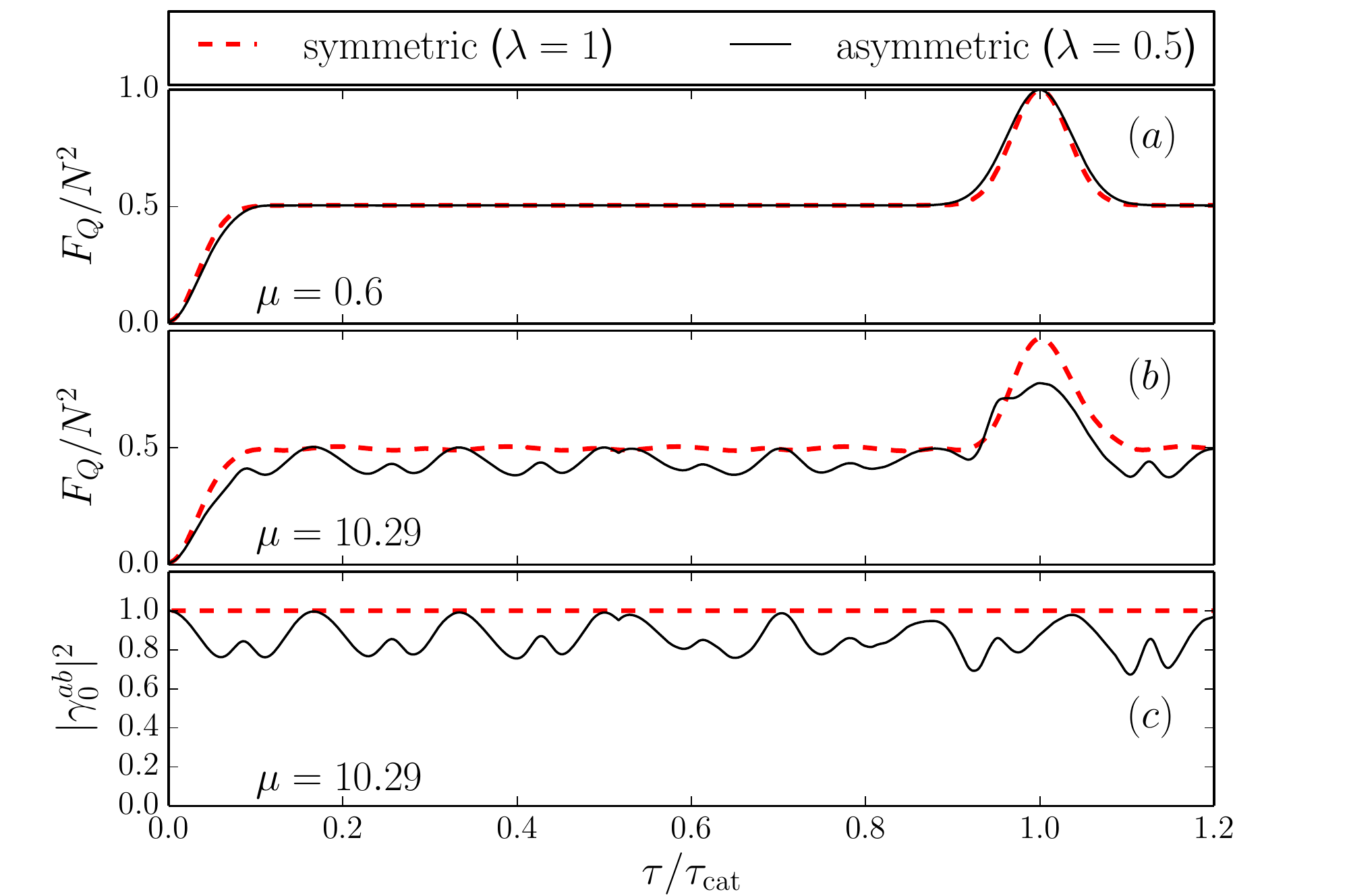}
\caption{For $N=100$ and $\kappa=0$ we compare the QFI of a condensate with symmetric interactions ($\lambda=1$) and an asymmetric condensate ($\lambda=0.5$) for different $\mu$ (\textbf{a}, \textbf{b}). Asymmetric condensates are far more susceptible to the deleterious effects of multi-mode dynamics. The cat times for the symmetric (asymmetric) condensates are $\tau_{\mathrm{cat}}=1570 (1814)$, $18.06 (20.33)$ for $\mu=0.6$, $10.29$, respectively. We also compare the magnitude of the overlap $\gamma^{ab}_{0}$ (\textbf{c}) for symmetric and asymmetric condensates.}
\label{fig:asymm}
\end{figure}

In condensates with symmetric interactions ($\lambda=1$) the maximal, zeroth-order overlap $\gamma^{ab}_0(0)=1$ due to symmetry and normalisation of the mode functions. In asymmetric condensates, even this term is reduced indicating that asymmetric condensates are poorly suited to the creation of spin-cat states. This intuition is confirmed in Figure \ref{fig:asymm}, which compares the QFI for symmetric interactions ($\lambda=1$) and asymmetric interactions ($\lambda=0.5$) with otherwise identical parameters. Although the two agree well in the single-mode regime [Figure \ref{fig:asymm}(a)], increasing $\mu$ diminishes the peak QFI of the asymmetric system compared to the symmetric one [Figure \ref{fig:asymm}(c)]. Asymmetric systems require a $\pi$-pulse roughly at $\tau_{\mathrm{cat}}/2$, the precise timing of this pulse is found by numerically optimising the peak QFI. In Figure \ref{fig:asymm} (c) we show the maximal zeroth-order overlap $\gamma^{ab}_0(0)$, which is always unity in symmetric systems, but is significantly reduced in the asymmetric simulation. This is in addition to the loss of QFI due to reduced $\gamma^{jk}_2$, indicating that asymmetric systems are more susceptible to the deleterious effects of multi-mode dynamics. There is however a speed-up to be gained over symmetric condensates [c.f. Eq.~\eqref{eq:tcatTF}], indicating that working with asymmetric condensates may still be desirable, especially if $\mu$ is close to $E_0$ where the multi-mode dynamics are less deleterious.

\section{Speed Limits on Cat Evolution Time} \label{sec:MMcats}

Having determined that multi-mode dynamic can have a deleterious effect on the peak QFI, we performed many simulations such as those depicted in Figures \ref{fig:catpeaks} and \ref{fig:asymm}. The results are collected in Figure \ref{fig:fqvsmu}. As condensates with asymmetric interactions always perform more poorly than symmetric ones for otherwise fixed parameters, we focus on symmetric systems ($\lambda=1$). In particular we collate the maximum QFI and the corresponding time $\tau_{\mathrm{cat}}$ as a function of chemical potential. The maximum QFI is compared to $F_Q=N^2/2$, which is approximately the QFI of a twin-Fock state (TFS) $|TFS \rangle = |N/2, N/2 \rangle$, which has $N/2$ atoms in both components. This QFI which is quickly reached under one-axis twisting (see for instance Figures \ref{fig:QFI_sm}, \ref{fig:catpeaks}) and serves as a threshold - if the maximum QFI is close to the TFS QFI there is little point bothering with the comparatively large $\tau_{\mathrm{cat}}$. 

The top row of Figure \ref{fig:fqvsmu} reveals that, as $\mu$ increases and moves the system away from the single-mode regime, there are certain values of $\mu/N$ that support the creation of states with QFI approaching $F_Q=N^2$. Crucially, this is true even in the presence of significant multi-mode dynamics. After scaling out the total number $N$, for a particular value of $\kappa$ these values of $\mu/N$ are predicted by studying the period, $T$, of the (maximally weighted) overlap $|\gamma_2^{jk}(m=0)|$, which we previously deduced is primary responsible for the decay in QFI. The bottom row of Figure \ref{fig:fqvsmu} shows the time at which the maximum QFI occurs (empty squares, not necessarily $t_\mathrm{{cat}}$), compared to the period of $|\gamma_2^{aa}(0)|$. Values of $\mu$ with $\tau_{\mathrm{cat}}$ [Eq.~\eqref{eq:tcatTF}, in Figure \ref{fig:fqvsmu} plotted as a solid black line] coinciding with an integer multiple of $T$ give rise to cat peaks significantly above the $N^2/2$ threshold (dashed magenta line). The take-home result of this plot is this: for a symmetric condensate far from the single-mode regime, given an $N$ and $\kappa$, it is not possible to use OAT to generate a state with QFI comparable to a spin-cat state faster than $T$, and $\mu$ should be chosen such that $t_{\mathrm{cat}}$ is an integer multiple of $T$. 


\section{Conclusion}

We have investigated the creation of spin-cat states from non-linear atomic interactions in a two-level 1D trapped Bose-Einstein condensate. Starting from a simple analytic treatment within the Thomas-Fermi approximation, we deduce that working in a multi-mode regime should produce a larger twisting rate, and thus faster evolution toward a spin-cat state in absolute terms. Using the method of Sinatra and Castin we find that multi-mode dynamics have a deleterious effect on the maximum QFI, as a result of poor overlap between the mode functions of the different spatial components, especially the second order overlaps $\gamma_2^jk$. In particular we find that condensates with asymmetric interactions are more susceptible to this overlap reduction than symmetric condensates. For this reason we focus on symmetric interactions, and find that even in the presence of significant multi-mode dynamics, maximum QFI close to $N^2$ is possible so long as $t_{\mathrm{cat}}=\pi/2\chi$ matches the period of $|\gamma_2^{jk}(m=0)|$ [Figure \ref{fig:fqvsmu}]. This time scale imposes a speed limit on the time taken for one-axis twisting to produce a state with QFI close to the Heisenberg limit.


\begin{acknowledgments}

The authors would also like to thank Joel Corney and Stuart Szigeti for invaluable discussion and feedback. This project has received funding from the European Union's Horizon 2020 research and innovation programme under the Marie Sklodowska-Curie grant agreement No 704672. 

\end{acknowledgments}

\appendix
\section{The Method of Sinatra and Castin} \label{sec:appA}

In this appendix we will present the method of Sinatra and Castin, and derive a number of useful results which we use throughout this paper.

\subsection{Example: Single component system}

For simplicity, we will first present the method as it applies to a single-component field $\hat{\psi}(\mathbf{r})$. In Section \ref{sec:twolevel} we present the generalisation to a two-component system represented by the fields $\hat{\psi}_a(\mathbf{r})$, $\hat{\psi}_b(\mathbf{r})$. The idea behind the method is to expand the state in the number basis, and then evolve \emph{each} number state within a single-mode (or Hartree-Fock) approximation. Quantitatively, this is done by defining a Bosonic annihilation operator
\begin{equation} \label{eq:aphi}
\hat{a}_{\phi} = \int d \mathbf{r} \phi^*(\mathbf{r}) \hat{\psi}(\mathbf{r}).
\end{equation}
which destroys a particle with mode function $\phi(\mathbf{r})$ (normalised to unity). From the cannonical commutation relations $\left[\hat{\psi}(\mathbf{r}), \hat{\psi}^\dagger (\mathbf{r}') \right] = \delta(\mathbf{r}-\mathbf{r}')$ it follows that
\begin{equation} \label{eq:commu}
\left[\hat{a}_{\phi(\mathbf{r})}, \hat{a}^\dagger_{\varphi(\mathbf{r}')} \right] = \int d \mathbf{r} \phi^*(\mathbf{r}) \varphi(\mathbf{r}) .
\end{equation}
It is possible to use the mode-annihilation operator Eq.~\eqref{eq:aphi} to construct a set of number-like states. To start with, the mode-annihilation operator Eq.~\eqref{eq:aphi} acts on the vacuum in the standard way,
\begin{equation} \label{eq:modevac}
\hat{a}_\phi |0 \rangle = 0 .
\end{equation}
Thus, we can define the basis states 
\begin{equation} \label{eq:kindofFock}
|n; \phi_n \rangle = \frac{\left( \hat{a}^\dagger_{\phi_n}\right)^n}{\sqrt{n!}} |0 \rangle ,
\end{equation}
which each have their own mode function $\phi_n(\mathbf{r})$. This is conceptually similar to the usual single-mode approximation, with the key difference that each number component has its own wavefunction (or mode function). Associating each number state with a unique mode function means that superpositions of these states
\begin{equation} \label{eq:SMstate}
|\psi(t) \rangle = \sum_{n=1}^N c_n |n; \phi_n(\mathbf{r}, t) \rangle 
\end{equation}
can exhibit spatial dynamics. The utility of this picture is that if written in this basis, $|\psi(t) \rangle$ can be evolved in time by simply evolving each mode function $\phi_n(\mathbf{r},t)$, i.e. in this picture the basis states are time \emph{dependent}. Essentially the full multi-mode problem is recast as a single mode problem with spatial dynamics accounted for separately by $\phi_n(\mathbf{r},t)$, which can significantly reduce the numerical difficulty of constructing and evolving the full state. Thus we have retained a multi-mode description of the system, within the approximation that \emph{each} number component is well described by a single wavefunction. As a counter-example, another $n$-particle state is
\begin{equation}
| \tilde{n} \rangle =  \frac{1}{\sqrt{n!}} \left( \hat{a}^\dagger_{\phi_{n-1}}\right)^{n-1} \hat{a}^\dagger_{\phi_1}  |0 \rangle ,
\end{equation}
which cannot be expressed in the form of Eq.~\eqref{eq:kindofFock} if $\phi_1 \neq \phi_{n-1}$.

Making use of Eq.~\eqref{eq:modevac} and applying commutator Eq.~\eqref{eq:commu} $m$ times with the identity
\begin{equation} \label{eq:commuidentity}
[\hat{A}, \hat{B}^n]=n\hat{B}^{n-1}[\hat{A}, \hat{B}], 
\end{equation} 
(which holds for any two operators $\hat{A}$, $\hat{B}$ that both commute with $[\hat{A}, \hat{B}]$), the overlap of the basis states is
\begin{equation} \label{eq:singlemodeoverlap}
\langle m; \phi_{m'}(\mathbf{r}') | n; \phi_{n'}(\mathbf{r}) \rangle = \delta_{m,n} \left( \int d \mathbf{r} \phi^*_{m'}(\mathbf{r}) \phi^*_{n'}(\mathbf{r}) \right)^m .
\end{equation}
Although the mode functions $\phi_n(\mathbf{r})$ are non-orthogonal, they are normalised to unity. Importantly, this implies that the states $|n; \phi_{n'}(\mathbf{r}) \rangle$ are orthogonal, and normalised so long as $m=m'$ and $n=n'$, which justifies the expansion Eq.~\eqref{eq:SMstate}. 

\subsection{Equations of motion for a single component field.} \label{sec:eqofmot}

Here we will derive equations of motion for the mode functions $\phi_n(\mathbf{r},t)$, and then discuss how these may be used to calculate expectation values. 

For brevity, in this section we will suppress the $n$ index on the mode functions $\phi_n(\mathbf{r},t)$. Consider a single component system evolving under the Hamiltonian 
\begin{align} \label{eq:BECHam2}
\hat{\mathcal{H}} = \int d \mathbf{r} \hat{\psi}^\dagger (\mathbf{r}) \hat{H}_0 \hat{\psi} (\mathbf{r}) + \frac{g}{2} \int d \mathbf{r} \hat{\psi}^\dagger (\mathbf{r})\hat{\psi}^\dagger (\mathbf{r}) \hat{\psi} (\mathbf{r}) \hat{\psi} (\mathbf{r}) ,
\end{align}
with single-particle Hamiltonian $\hat{H}_0$. Assuming the coefficients $c_n$ are stationary, then the dynamics are entirely governed by the basis states $|n,\phi(\mathbf{r}, t) \rangle$, which evolve under the Schr{\"o}dinger equation

\begin{equation}
i\hbar \frac{\partial}{\partial t} |n; \phi(\mathbf{r},t) \rangle = \hat{\mathcal{H}}|n; \phi(\mathbf{r},t) \rangle .
\end{equation} 
Using the definition Eq.~\eqref{eq:kindofFock} with Eq.~\eqref{eq:aphi}, the left-hand side (LHS) of the Schr{\"o}dinger equation is 
\begin{equation}
i\hbar \frac{\partial}{\partial t} |n; \phi(\mathbf{r},t) \rangle = \frac{i \hbar n}{\sqrt{n!}} \left( \hat{a}^\dagger_\phi \right)^{n-1} \left(\int d \mathbf{r} \frac{\partial \phi(\mathbf{r},t)}{\partial t} \hat{\psi}^\dagger(\mathbf{r}) \right) |0 \rangle .
\end{equation}
However the right-hand side (RHS) is a little more effort, so we split the RHS into the single-particle term and the interaction term. To proceed, the commutators
\begin{align} 
\label{eq:psiadagcommu}
\left[ \hat{\psi}(\mathbf{r}), \hat{a}^\dagger_{\phi(\mathbf{r}',t)} \right]&=\phi(\mathbf{r},t) \\
\left[ \hat{a}_{\phi(\mathbf{r},t)}, \hat{\psi}^\dagger (\mathbf{r}') \right]&=\phi^*(\mathbf{r}',t)
\end{align}
will be needed, along with the commutator identity Eq.~\eqref{eq:commuidentity}. The single particle term is

\begin{align}
& \int d\mathbf{r} \hat{\psi}^\dagger(\mathbf{r})\hat{H}_0 \hat{\psi}(\mathbf{r}) |n; \phi(\mathbf{r}',t) \rangle \\
&= \frac{n}{\sqrt{n!}}\int d\mathbf{r} \hat{\psi}^\dagger(\mathbf{r}) \hat{H}_0 \phi(\mathbf{r},t) \left(\hat{a}^\dagger_{\phi(\mathbf{r}',t)} \right)^{n-1} |0 \rangle , \nonumber
\end{align}
and the interaction term is
\begin{align}
& \frac{g}{2} \int d \mathbf{r} \hat{\psi}^\dagger (\mathbf{r})\hat{\psi}^\dagger (\mathbf{r}) \hat{\psi} (\mathbf{r}) \hat{\psi} (\mathbf{r}) |n; \phi(\mathbf{r}',t) \rangle\\ \nonumber
& = \frac{g n(n-1)}{2\sqrt{n!}} \int d\mathbf{r} \hat{\psi}^\dagger(\mathbf{r}) \hat{\psi}^\dagger(\mathbf{r}) \phi(\mathbf{r},t)^2 \left(\hat{a}^\dagger_{\phi(\mathbf{r}',t)} \right)^{n-2} |0 \rangle \nonumber .
\end{align}
Equating the LHS and the RHS of the Schr{\"o}dinger equation, and multiplying through on the left by the inverse of $\hat{a}^\dagger_\phi$ $n-2$ times, we obtain 
\begin{widetext}
\begin{align} \label{eq:notGPEyet2}
&\Bigg( \hat{a}^\dagger_{\phi(\mathbf{r},t)}  \int d\mathbf{r} \hat{\psi}^\dagger (\mathbf{r}) \bigg( i\hbar \frac{\partial \phi(\mathbf{r},t)}{\partial t}  -\hat{H}_0 \phi(\mathbf{r},t) \bigg) - \frac{g}{2} (n-1) \int d\mathbf{r} \hat{\psi}^\dagger(\mathbf{r}) \hat{\psi}^\dagger(\mathbf{r}) \phi(\mathbf{r},t)^2 \Bigg) |0 \rangle \\ 
&=  \Bigg( \hat{a}_{\phi(\mathbf{r},t)} \hat{a}^\dagger_{\phi(\mathbf{r},t)}  \int d\mathbf{r} \hat{\psi}^\dagger (\mathbf{r}) \bigg( i\hbar \frac{\partial \phi(\mathbf{r},t)}{\partial t}  -\hat{H}_0 \phi(\mathbf{r},t) \bigg) - g (n-1) \int d\mathbf{r} \hat{\psi}^\dagger(\mathbf{r})|\phi(\mathbf{r},t)|^2 \phi(\mathbf{r},t) \Bigg) |0 \rangle \\
&=0 ,
\end{align}
\end{widetext}
where in the second line we have multiplied through by $\hat{a}_\phi$ on the left. Finally, normally ordering $\hat{a}_\phi \hat{a}^\dagger_\phi$ gives
\begin{equation} \label{eq:stillnotGPE}
\left(\hat{a}^\dagger_\phi f(t) + \int d \mathbf{r} \hat{\psi}^\dagger ( \mathbf{r}) g(\mathbf{r},t) \right)| 0 \rangle = 0 ,
\end{equation}
where we have defined:
\begin{align}
\label{eq:ft}
f(t) &= \int d \mathbf{r} \phi^*(\mathbf{r},t) \left(i \hbar \frac{\partial \phi(\mathbf{r},t)}{\partial t} - \hat{H}_0 \phi(\mathbf{r},t) \right) , \\ 
g(\mathbf{r},t) & = i \hbar \frac{\partial \phi(\mathbf{r},t)}{\partial t} - \hat{H}_0 \phi(\mathbf{r},t) - g(n-1) |\phi(\mathbf{r},t)|^2  \phi(\mathbf{r},t) . 
\end{align}
Eq.~\eqref{eq:stillnotGPE} implies,
\begin{align}
\int d \mathbf{r} \hat{\psi}^\dagger ( \mathbf{r}) g(\mathbf{r},t) &=  -\hat{a}^\dagger_\phi f(t) \\ 
&= - \int d \mathbf{r} \hat{\psi}^\dagger(\mathbf{r}) f(t) \phi(\mathbf{r},t) ,
\end{align}
where the second line follows from the definition of $\hat{a}_\phi$ [Eq.~\eqref{eq:aphi}]. Equating the integrands we can conclude that $g(\mathbf{r},t) = -f(t) \phi(\mathbf{r},t)$, which can be re-written as
\begin{equation} \label{eq:almostGPE}
i \hbar \frac{\partial \phi_n(\mathbf{r},t)}{\partial t} = \left(\hat{H}_0 - f_n(t) + g(n-1) |\phi_n(\mathbf{r},t)|^2 \right) \phi_n(\mathbf{r},t) .
\end{equation}
Notice we have re-labelled $f(t)$ and the mode functions with the $n$ index. This equation is strikingly similar to the time-dependent Gross-Pitaevskii equation (GPE), but with a dynamic offset to the energy given by $f_n(t)$. Substituting Eq.~\eqref{eq:almostGPE} into Eq.~\eqref{eq:ft} reveals that $f_n(t)$ depends only on the non-linear term,
\begin{equation} \label{eq:ft2}
f_n(t) = \frac{g}{2} (n-1) \int d\mathbf{r} |\phi_n(\mathbf{r},t)|^4 . 
\end{equation}

In principle, Eq.~\eqref{eq:almostGPE} fully determines the dynamics of $\phi_n(\mathbf{r},t)$, i.e. one must solve the equation of motion for each $\phi_n(\mathbf{r},t)$, computing $f_n(t)$ at each time step via Eq.~\eqref{eq:ft2}. However it is tidier to define the rotated field 
\begin{equation}
\tilde{\phi}_n(\mathbf{r},t) = \phi_n(\mathbf{r},t) e^{-i A_n(t)/\hbar n} ,
\end{equation}
with 
\begin{equation}
A_n(t) = -n \int_0^t f_n(t') dt'.
\end{equation}
To summarise, the time-evolved basis state is
\begin{equation} \label{eq:basisevolve}
|n; \phi_n(\mathbf{r},t) \rangle = \frac{e^{-i A_n(t)/\hbar}}{\sqrt{n!}} \left( \int d\mathbf{r} \tilde{\phi}_n(\mathbf{r},t) \hat{\psi}^\dagger(\mathbf{r}) \right)^n |0 \rangle ,
\end{equation}
where $\tilde{\phi}_n(\mathbf{r},t)$ evolve under a set of GPE-like equations
\begin{equation} \label{eq:CSGPE}
i \hbar \frac{\partial \tilde{\phi}_n(\mathbf{r},t)}{\partial t} = \left(\hat{H}_0 + g(n-1) |\tilde{\phi}_n(\mathbf{r},t)|^2 \right) \tilde{\phi}_n(\mathbf{r},t) ,
\end{equation}
and $A_n(t)$ are computed by integrating
\begin{equation}\label{eq:CSAt}
\frac{dA_n(t)}{dt} =  - \frac{g}{2}n(n-1) \int d\mathbf{r} |\tilde{\phi}_n(\mathbf{r},t)|^4 .
\end{equation}

For a single field containing $N$ atoms, the full state $|\psi(t) \rangle$ can be constructed by solving $N$ coupled differential equations Eq.~\eqref{eq:CSGPE} and Eq.~\eqref{eq:CSAt}. However, this may still be a challenging numerical task if $N$ is large, particularly in higher numbers of spatial dimensions. Fortunately, for some states it may be sufficient to consider a subset of the full system. Additionally, in Ref. \cite{Li2009} the authors present a number of approximation methods which are less computationally demanding, however these methods are not employed in this work and as such they will not be discussed any further. Here, we will calculate some observables in terms of $\tilde{\phi}_n(\mathbf{r},t)$ and $A_n(t)$. For simplicity, we drop the tilde notation, with the understanding that $\phi_n(\mathbf{r},t)$ is the solution to Eqs.~\eqref{eq:CSGPE} rather than Eq.~\eqref{eq:almostGPE} with Eq.~\eqref{eq:ft2}. 

\subsection{Generalisation to two fields} \label{sec:twolevel}

In the previous section we presented the case of a single field mainly for pedagogical reasons. In this section we present the formalism as it applies two fields $\hat{\psi}_a$, $\hat{\psi}_b$. In analogy with Eq.~\eqref{eq:kindofFock}, the relevant (dynamic) basis states are 
\begin{align} \label{eq:twomodebasis}
& |\mathbf{n}; \phi_{a,\mathbf{n}}(\mathbf{r}, t), \phi_{b, \mathbf{n}}(\mathbf{r}, t) \rangle \\
& = e^{-i A_{\mathbf{n}}(t)/\hbar}\frac{\left(\hat{a}^\dagger_{\phi_{a,\mathbf{n}}} \right)^{n_a} }{\sqrt{n_a!}}\frac{\left(\hat{b}^\dagger_{\phi_{b,\mathbf{n}}} \right)^{n_b}}{\sqrt{n_b!}} |0\rangle ,
\end{align}
with $\mathbf{n}=(n_a, n_b)$, and
\begin{align}
\hat{a}_{\phi_{a,\mathbf{n}}} &= \int d \mathbf{r} \phi_{a,\mathbf{n}}^*(\mathbf{r},t) \hat{\psi}_a(\mathbf{r}) \\
\hat{b}_{\phi_{b,\mathbf{n}}} &= \int d \mathbf{r} \phi_{b,\mathbf{n}}^*(\mathbf{r},t) \hat{\psi}_b(\mathbf{r}) .
\end{align}

Consider the two-component Hamiltonian
\begin{align}
\mathcal{H} &= \sum_{j=a,b} \hat{\psi}_j^\dagger(\mathbf{r}) \hat{H}_0 \hat{\psi}_j(\mathbf{r}) + \\
& + \sum_{j,k=a,b} \frac{g_{jk}}{2} \int d \mathbf{r}  \hat{\psi}_j^\dagger(\mathbf{r})  \hat{\psi}_k^\dagger(\mathbf{r})  \hat{\psi}_j(\mathbf{r}) \hat{\psi}_k(\mathbf{r}) .
\end{align}
Following a similar procedure to the previous Section, one can derive the equations of motion for the mode functions $\phi_{a,\mathbf{n}}(\mathbf{r},t)$, $\phi_{b,\mathbf{n}}(\mathbf{r},t)$ and phase factor $A_{\mathbf{n}}(t)$ (suppressing the $\mathbf{r}$ and $t$ dependence),
\begin{align} \label{eq:CSeqofmot2a}
i \hbar \frac{\partial  \phi_{a,\mathbf{n}}}{\partial t} &= \left( \hat{H}_0 + g_{aa}(n_a-1)|\phi_{a, \mathbf{n}}|^2 + g_{ab} n_b |\phi_{b, \mathbf{n}}|^2 \right) \phi_{a, \mathbf{n}} \\
\nonumber \\
i \hbar \frac{\partial  \phi_{b,\mathbf{n}}}{\partial t} &= \left( \hat{H}_0 + g_{bb}(n_b-1)|\phi_{a, \mathbf{n}}|^2 + g_{ab} n_a |\phi_{a, \mathbf{n}}|^2 \right) \phi_{b, \mathbf{n}} \\
\nonumber \\\label{eq:CSeqofmot2b}
\frac{dA_{\mathbf{n}}}{dt} &= - \sum_{j=a,b} \frac{g_{jj}}{2}n_j(n_j-1) \int d\mathbf{r} |\phi_{j, \mathbf{n}}|^4 - \\
& - g_{ab} n_a n_b \int d\mathbf{r} |\phi_{a, \mathbf{n}}|^2 |\phi_{b, \mathbf{n}}|^2 . \nonumber
\end{align}

Once these equations have been solved, expectation values of normally ordered moments (such as those presented in Section \ref{sec:twomodeobservables}) can be evaluated making use of the following identities,
%
\begin{align} \label{eq:psifocka}
& \hat{\psi}_a(\mathbf{r}) |\mathbf{n}; \phi_{a, \mathbf{n}'}(\mathbf{r}', t), \phi_{b, \mathbf{n}'}(\mathbf{r}', t) \rangle \nonumber \\
&= \sqrt{n_a} \phi_{a, \mathbf{n}'}(\mathbf{r}, t) |(n_a-1, n_b); \phi_{a, \mathbf{n}'}(\mathbf{r}', t),\phi_{b, \mathbf{n}'}(\mathbf{r}', t) \rangle \\ \label{eq:psifockb}
& \hat{\psi}_b(\mathbf{r}) |\mathbf{n}; \phi_{a, \mathbf{n}'}(\mathbf{r}', t), \phi_{b, \mathbf{n}'}(\mathbf{r}', t) \rangle \nonumber \\
& = \sqrt{n_b} \phi_{b, \mathbf{n}'}(\mathbf{r}, t) |(n_a, n_b-1); \phi_{a, \mathbf{n}'}(\mathbf{r}', t), \phi_{b, \mathbf{n}'}(\mathbf{r}', t) \rangle
\end{align}
with the overlap 
%
\begin{align} \label{eq:SU2LSoverlap}
& \langle \mathbf{m}; \phi_{a, \mathbf{m}'}(\mathbf{r}', t), \phi_{b, \mathbf{m}'}(\mathbf{r}', t) |\mathbf{n}; \phi_{a, \mathbf{n}'}(\mathbf{r}, t), \phi_{a, \mathbf{n}'}(\mathbf{r}, t) \rangle \\
& =  \left(\int d \mathbf{r} \phi_{a,\mathbf{m}'}^*(\mathbf{r}, t) \phi_{a,\mathbf{n}'}(\mathbf{r}, t) \right)^{n_a}  \times \nonumber \\
& \times \left(\int d \mathbf{r} \phi_{b,\mathbf{m}'}^*(\mathbf{r}, t) \phi_{b,\mathbf{n}'}(\mathbf{r}, t) \right)^{n_b} e^{i[A_{\mathbf{m}'}(t)-A_{\mathbf{n}'}(t)]/\hbar} \delta_{\mathbf{m},\mathbf{n}} \nonumber .
\end{align}
Note that $\delta_{\mathbf{m}, \mathbf{n}} = \delta_{m_a, n_a} \delta_{m_b, n_b}$ and also that typically $\mathbf{n}$, $\mathbf{n}'$ are not independent (for instance $n_a'=n_a \pm 1$, and likewise for $\mathbf{m}$, $\mathbf{m}'$).

\subsection{SU(2) expectation values} \label{sec:twomodeobservables}

In the main text we are interested in employing this formalism to calculate the expectation values of observables associated with the SU(2) algebra. If we impose the constraint $n_a+n_b=N$, then all two-level basis states Eq.~\eqref{eq:twomodebasis} have conserved total number $N$, which can now be labelled with a \emph{single} index $m=(n_a-n_b)/2$. This is the notation used in the main text [Eq.~\eqref{eq:chap7state}]. These states are the multi-mode analogue of Dicke states. We note that for any SU(2) state summing over $m$ is equivalent to summing over $n_a$, but we continue to use the latter for notation for clarity.

We study the time-evolution of coherent-spin states of $N$ atoms, i.e.
\begin{align} \label{eq:CSS_castinsinatra}
|\psi(t) \rangle &= \frac{1}{\sqrt{N!}} \left(c_a \hat{a}^\dagger_{\phi_a} + c_b \hat{b}^\dagger_{\phi_b} \right)^N |0 \rangle \\
&= \sum_{n_a=0}^{N} \sqrt{\frac{N!}{n_a!n_b!}} (c_a)^{n_a} (c_b)^{n_b}  \times \\
& \times |m; \phi_{a,m}(\mathbf{r}, t), \phi_{b,m}(\mathbf{r}, t) \rangle . \nonumber
\end{align}
The coefficients $c_a$, $c_b$ describe the initial coherent-spin state with $|c_a|^2+|c_b|^2=1$, and initial condition $\phi_a=\phi_b$. 

Observables that commute with the total number are conserved, and can be simply evaluated with respect to the initial coherent-spin state Eq.~\eqref{eq:CSS_castinsinatra}, for instance $\langle \hat{N}_a \rangle = N/2$, $\langle \hat{N}_a^2 \rangle=N^2/4$, etc. In general expectation values of normally-ordered observables can be evaluated by making use of Eq.~\eqref{eq:psifocka} and Eq.~\eqref{eq:psifockb}, with the overlap Eq.~\eqref{eq:SU2LSoverlap}. In terms of the mode-overlaps
\begin{equation}
\gamma^{jk}_{m^\prime}(m, t) = \int d\mathbf{r} \phi_{j,m}(\mathbf{r},t) \phi_{k, m-m^\prime}^*(\mathbf{r},t) ,
\end{equation}
we have 
%

\begin{widetext}
\begin{align} \label{eq:CSJp}
\langle \hat{J}_+ \rangle &= \int d \mathbf{r} \langle \hat{\psi}_a(\mathbf{r}) \hat{\psi}^\dagger_b(\mathbf{r}) \rangle \\
&= \sum_{n_a=1}^{N} \frac{N!}{(n_a-1)! n_b!} |c_a|^{2(n_a-1)} |c_b|^{2n_b} c_b^* c_a  e^{i [A_{m-1}(t)-A_m(t)]/\hbar} \gamma^{ab}_1(m,t) \left[\gamma^{aa}_1(m,t) \right]^{n_a-1} \left[\gamma^{bb}_1(m,t) \right]^{n_b} .\nonumber 
\end{align}
\end{widetext}

Up to conjugates and ordering, there are four unique moments which are needed to study one-axis twisting. The relevant higher order moments are (see also Ref.\cite{Li2009}, Appendix A):
\begin{widetext}
\begin{align}
\langle \hat{J}_+ \hat{J}_- \rangle - \langle \hat{N}_b \rangle &= \int  \int d \mathbf{r} d \mathbf{r}'  \langle \hat{\psi}^\dagger_b(\mathbf{r}) \hat{\psi}^\dagger_a(\mathbf{r}') \hat{\psi}_a(\mathbf{r}) \hat{\psi}_b(\mathbf{r}') \rangle \\
&= \sum_{n_a=1}^{N-1} \frac{N!}{(n_a-1)! (n_b-1)!} |c_a|^{2n_a} |c_b|^{2n_b} \left| \gamma^{ab}_0(m,t)\right|^2  \nonumber 
\end{align}
%
\begin{align}
\langle \hat{J}_+ \hat{J}_+ \rangle &= \int  \int d \mathbf{r} d \mathbf{r}' \langle \hat{\psi}^\dagger_b(\mathbf{r}) \hat{\psi}^\dagger_b(\mathbf{r}') \hat{\psi}_a(\mathbf{r}) \hat{\psi}_a(\mathbf{r}') \rangle \\
&=\sum_{n_a=2}^{N} \frac{N!}{(n_a-2)! n_b!} |c_a|^{2(n_a-2)} |c_b|^{2n_b} (c_b^*)^2 c_a^2 e^{i [A_{m-2}-A_m]/\hbar}  \left[ \gamma^{ab}_2(m,t) \right]^2\times \left[ \gamma^{aa}_2(m,t) \right]^{n_a-2} \left[\gamma^{bb}_2(m,t) \right]^{n_b}  \nonumber 
\end{align}
%
\begin{align}
\langle \hat{N}_b \hat{J}_+ \rangle - \langle \hat{J}_+ \rangle &= \int d \mathbf{r} d \mathbf{r}' \langle \hat{\psi}^\dagger_b(\mathbf{r}) \hat{\psi}^\dagger_b(\mathbf{r}') \hat{\psi}_b(\mathbf{r}) \hat{\psi}_a(\mathbf{r}') \rangle \\
&= \sum_{n_a=1}^{N-1} \frac{N!}{(n_a-1)! (n_b-1)!} |c_a|^{2(n_a-1)} |c_b|^{2n_b} c_b^* c_a  e^{i [A_{m-1}-A_m]/\hbar} \gamma^{ab}_1(m,t) \left[\gamma^{aa}_1(m,t) \right]^{n_a-1} \left[ \gamma^{bb}_1(m,t) \right]^{n_b} \nonumber 
\end{align}
%
\begin{align} \label{eq:lastmoment}
\langle \hat{N}_a \hat{J}_- \rangle - \langle \hat{J}_- \rangle &= \langle \hat{\psi}^\dagger_a(\mathbf{r}) \hat{\psi}^\dagger_a(\mathbf{r}') \hat{\psi}_a(\mathbf{r}) \hat{\psi}_b(\mathbf{r}') \rangle \\
&= \sum_{n_a=1}^{N} \frac{N!}{(n_a-1)! (n_b-1)!} |c_a|^{2n_a} |c_b|^{2(n_b-1)} c_a^* c_b e^{i [A_{m+1}-A_m]/\hbar} \gamma^{ba}_{-1}(m,t) \left[\gamma^{aa}_{-1}(m,t) \right]^{n_a} \left[\gamma^{bb}_{-1}(m,t) \right]^{n_b-1} \nonumber  .
\end{align}
\end{widetext}

\bibliographystyle{apsrev4-1}
\bibliography{cat_bib}

\end{document}